\documentstyle[aps,epsfig,multicol,color,times,eqsecnum]{revtex}
\newcommand{\infig}[2]{\begin{center}{\epsfig{file=#2,width=#1}}\end{center}} 
\newcommand{\Beginfigure}{\vskip 0pt\noindent}
\newcommand{\Endfigure}{\vskip 0pt\noindent}
\newcommand{\Caption}[1]{\par\noindent{\small #1} \par}

\newcommand{\Endrule}{\vskip 3pt\noindent\hrule width 8.6cm\vskip 3pt}
\newcommand{\Beginrule}{\vskip 3pt\noindent\hbox{%
\vbox{\hbox to 9cm{\hfill}}\vbox{\hrule width 9cm}} \vskip 3pt}
\long\def\comment#1{\vskip 2mm\noindent\fbox{%
\vbox{\parindent=0cm{\small\em #1}}}\vskip 2mm}

\newcommand{\Startsmall}{}
\newcommand{\Endsmall}{}
\newcommand{\StartTwoColumn}{\begin{multicols}{2}}
\newcommand{\EndTwoColumn}{\end{multicols}}
\renewcommand{\narrowtext}{\Beginrule\begin{multicols}{2}\par\noindent}
\renewcommand{\widetext}{\end{multicols}\Endrule}
\def\Input#1{\input{#1}}
\newcommand{\Label}[1]{\label{#1}}
\newcommand{\rw}[1]{}
\def\DRAFT{%
\renewcommand{\Label}[1]{\label{##1}
{\hbox to 0cm{\textcolor{green}{\hss\em ##1\quad}}}}
\renewcommand{\rw}[1]{\vskip 10pt%
\noindent{\framebox{\textcolor{red}{New Material Needed}}}%
\par\noindent{\textcolor{red}{\em ##1}}\vskip 10pt}
\def\Input##1{\include{##1}}}
\newcommand{\UU}{Uehling-Uhlenbeck }

\begin{document}

\title{Quantum Kinetic Theory V: Quantum kinetic master equation for 
mutual interaction of condensate and noncondensate}
\author{C.W.~Gardiner$^{1}$ and P.~Zoller$^2$}
\address{$^1$ Physics Department, Victoria University, Wellington, New Zealand}
\address{$^2$ Insitut f{\"u}r Theoretische Physik,
Universit{\"a}t Innsbruck, 6020 Innsbruck, Austria}

\maketitle

\begin{abstract}A detailed quantum kinetic master equation is developed which 
couples the kinetics of a trapped condensate to the vapor of non-condensed 
particles.  This generalizes previous work
which treated the vapor as being undepleted. 
\end{abstract}

\pacs{PACS Nos. }
\long\def\comment#1{\vskip 2mm\noindent\fbox{%
\vbox{\parindent=0cm{\small\em #1}}}\vskip 2mm}

\StartTwoColumn
\section{Introduction}
Until the experimental realization of a Bose-Einstein condensate of 
magnetically trapped Alkali atoms \cite{JILA,Rice,MIT} in 
1995 (which has now become a widely available technology \cite{condensates})
the theory of Bose-Einstein condensation was dominated by the 
desire to understand the behavior of superfluid liquid helium, of 
which it was assumed that the weakly interacting degenerate Bose gas 
was a simplified and admittedly inadequate model.  
The central tools for the description of BEC were the Bogoliubov 
theory of superfluidity and modifications of it 
\cite{BOG,modBOG,trueBog}, the Gross-Pitaevskii equation
\cite{Gross-Pitaevskii} and techniques based on the Green's function 
formalism of many body theory \cite{Green}.  This philosophy is 
excellently presented in Griffin's book \cite{Griffin-QuLiquids}.

The prospect of actually producing a degenerate Bose gas gave rise to 
a number of investigations \cite{KEGrowth} based largely on a 
combination of kinetic theory as described by the Uehling-Uhlenbeck 
equation \cite{UU} and the the Gross-Pitaevskii equation.  However, 
nearly all of these investigations were from the point of view 
of macroscopic statistical mechanics, in which the BEC was seen as a 
sample of an extended fluid, rather than the tightly trapped 
condensate, which behaves more like a very large atom or molecule than 
a droplet of macroscopic fluid.

Our own program of quantum kinetic theory has been formulated in such 
a way as to take advantage the simplifications which can be made for a 
tightly trapped condensate.  It has been partially presented in our 
first four papers on the subject \cite{QK}, as well as in two papers 
on the growth of a Bose-Einstein condensates 
\cite{BosGro,NewestBosGro}.  However, all of the previous papers 
covered only special situations.  In QKI we treated only spatially 
homogeneous systems with at most a small amount of condensate; QKII 
dealt with stochastic equations as a model of the initiation of the 
condensate, QKIII and QKIV considered only the case where the majority 
of the atoms provided a bath of atoms with energy above a certain 
value (called $E_R$) of fixed chemical potential $\mu$ and temperature 
$T$ for a relatively small number of lower energy atoms, called the 
condensate band, which included the condesnsate level itself as well 
as the levels with energy less than $E_R$.

In this paper we will present what we see as a comprehensive and 
reasonably practical description of the subject, in which the various 
restrictions in the previous papers are largely eliminated.  The 
description which results is best described in terms of a system 
composed of a condensate and a vapor in interaction with each other.  
The vapor is described by a quantum kinetic master equation of the 
form introduced in QKI, but appropriately adapted to take account of 
the presence of a trap.  For nearly all situations, this is equivalent 
to a quantum Boltzmann equation of the Uehling-Uhlenbeck \cite{UU} 
form.  The condensate is described by a fully quantum mechanical 
Hamiltonian, which for many purposes can be treated by a Bogoliubov 
approximation, but truncated to include only those excitations not 
already included in the vapor.  The interaction between vapor and 
condensate consists of two parts; a Hamiltonian {\em mean field} term, 
and a {\em master equation} term which describes the transfer of 
energy and particles between vapor and condensate.  The largest part 
of the mean field term can be estimated, and included by defining {\em 
effective potentials} for the condensate and the vapor.  The 
computation of these effective potentials involves a procedure very 
similar to that of the Hartree-Fock-Bogoliubov \cite{HFB} method.  
However these effective potentials are used only to give a basis set 
of eigenfunctions, in terms of which we can describe the kinetics; the 
resulting kinetic equations give rise to corrections to the 
corresponding eigenvalues---our use of a {\em basis} of eigenfunctions 
similar to those given by the Hartree-Fock-Bogoliubov method does not 
mean our method is the same as the that method, since there is a 
residual term remaining after the majority of the mean field term has 
been included in the effective potentials, and this is of a similar 
order of magnitude to the master equation term.

This paper is concerned mainly with methodology and reasonably careful 
derivations, rather than with directly applicable results.  However, 
it provides a justification for the heuristic description of the 
theory of condensate growth presented in \cite{NewestBosGro}, and does 
provide a way in which hydrodynamics and quantum kinetics can be 
connected together.  There are, of course alternative approaches to 
the general issue of combining kinetic theory and quantum mechanics, 
which have different emphases.  For example Stoof \cite{StoofFPE} has 
chosen to develop an approach which avoids the use of the 
pseudopotential method, which he has shown will fail very near the 
critical point; Walser {\em et al.} \cite{WalserKinetic} have 
developed a kinetic approach somewaht more related to ours; while 
Zaremba {\em et al.} \cite{ZarembaGriffin} have developed a two fluid hydrodynamic 
description.  The major criterion for choosing a particular approach 
will ultimately be the degree of usefulness in explaining experiments, 
and its ease of use---but we will leave this judgment on this to the 
BEC community.

In the remainder of this introduction we will summarize the philosophy 
and scope of our methodology, and to emphasize the most significant results.

\subsection{The field theory of Bose particles}
We want to consider the Bose atoms to be described by a second-quantized 
field, in the pseudopotential approximation; that is, we write
\begin{eqnarray}\Label{In1}
H=H_{\rm kin}+H_I + H_T,
\end{eqnarray}
where
\begin{eqnarray}\Label{In2}
H_{\rm kin}&=&\int d^3{\bf x}\,\psi ^{\dagger }({\bf x})
\left(- {\hbar ^2\over2m}\nabla ^2\right) \psi ({\bf x})   ,
\\ \Label{In3}
H_I&=&{1\over 2}\int d^3{\bf x}\int d^3{\bf x}'\psi ^{\dagger }({\bf x})
\psi ^{\dagger }( {\bf x}') u( {\bf x}-{\bf x}') 
\psi ( {\bf x}') \psi ( {\bf x}) . 
\nonumber \\  
\end{eqnarray}
and the term $ H_T$ arises from a trapping potential as 
\begin{eqnarray}\Label{In4}
H_T =  \int d^3 {\bf x}\,V_T({\bf x)}\psi^\dagger({\bf x})\psi({\bf x}).
\end{eqnarray}
In this paper we will not go deeply into the approximations which allow us to 
use the pseudopotential form of the Hamiltonian, in which
\begin{eqnarray}\Label{In5}
u({\bf x}'-{\bf x})
 \to  u\delta({\bf x}'-{\bf x}) 
= {4\pi a\over m}\delta({\bf x}'-{\bf x}),
\end{eqnarray}
which is thoroughly investigated in many works \cite{pseudopotential}.  
Suffice it to say only that this is an approximation whose validity is 
assured provided the wavelengths of the atoms are very much longer 
than the range of the potential $ u({\bf x}'-{\bf x})$, and provided 
that we are not working too close to the critical point.  Thus, at 
typical temperatures and densities currently used, this approximation 
is valid either when there is a significant amount of condensate, or 
when there is no condensate.  The essence of the derivation of this 
pseudopotential approximation is really an adiabatic elimination of 
the fast motion of the atomic wavefunctions during the very small 
proportion of time during which they are actually undergoing 
collisions.  Thus the result is effectively a Hamiltonian which is 
coarse-grained in time, and thus, because of the dispersion relation, 
also in space.  The details of this philosophy are available in a 
number of references \cite{pseudopotential} Having made this 
approximation, we know we are then justified in using an approach 
which is essentially perturbative.

\subsection{Thermalization assumptions}
It is evident in experiments that a typical condensate in a trap is 
accompanied by a thermal vapor cloud.  The condensate consists of a 
large number of particles, all of which share the same wavefunction, 
while the vapor is essentially completely thermalized---it thus 
consists of particles which do not share wavefunctions with each 
other, and indeed most wavefunctions are not occupied by any particles 
at all.  The vapor consists mainly of the higher energy particles, but 
there is a transition from vapor to condensate which,
althought not sharp, necessitates  the drawing of a definited boundary between 
them.

\Beginfigure
\infig{8.5cm}{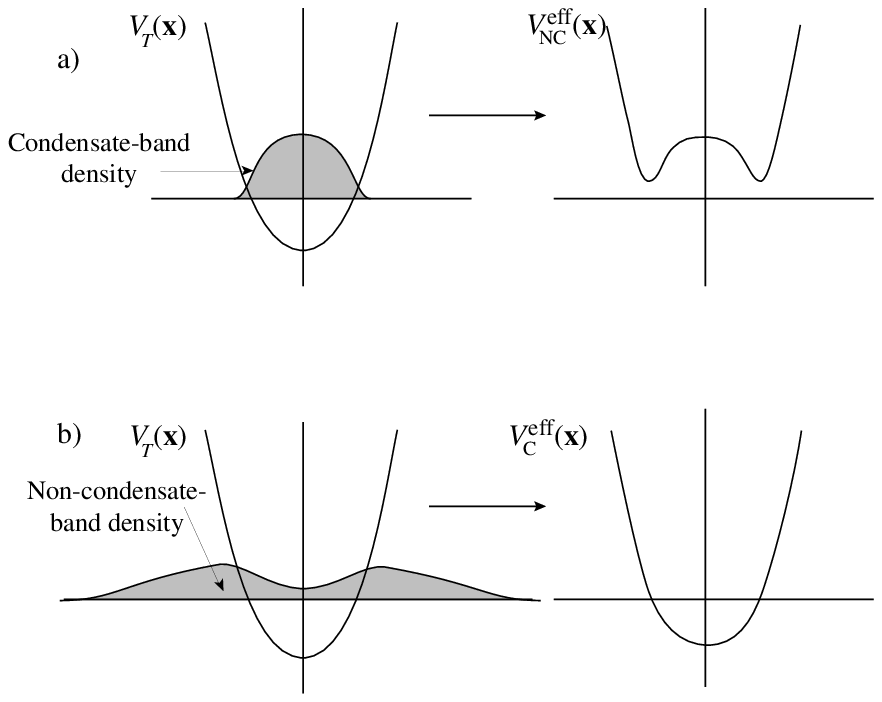} \Caption{Fig.~1: Representation of the 
modification of the trapping potential for a) the non-condensate band 
and b) the condensate band due to mean-field effects}
\Endfigure

\subsubsection{Condensate and non-condensate bands in the previous papers}
The concept of condensate and non-condensate bands was introduced by 
us in QKIII and \cite{BosGro}.  In these earlier works we 
chose to make the distinction between these in terms of the effect of 
the presence of a condensate on the trap energy levels.  This meant 
that the non-condensate band was taken to consist of those energy 
levels above a certain value, $E_R$, which were treated as being 
essentially unaffected by the condensate; the condensate band 
consisted of all levels with energies $E<E_R$, and these were taken to 
have wavefunctions and energies which were given by the Bogoliubov 
method.

In this paper we move to a more useful way of treating the two bands.  
In QKI we developed a wavelet description of weakly condensed 
systems, which allowed for a fully quantum-mechanical description of 
the thermalization, but this was derived  only for the case of free 
particles.  It was the absence of a good wavelet description of 
trapped particles that led us to consider in QKIII only the 
approximation that the non-condensate band could be treated as being 
like a bath of thermal particles with a fixed chemical potential, 
$\mu$ and a fixed temperature, $T$.  Attention was then concentrated 
on the derivation QKIII and successful application 
\cite{BosGro,QKIV,NewestBosGro} of a master equation for the 
condensate band.

\subsubsection{Condensate and non-condensate bands in this paper}
\Label{I.bands}
In this paper we have worked out how the wavelet description can be 
modified in the presence of a trapping potential, and have been able 
to give a full set of coupled equations connecting the two bands.  The 
distinction between the two bands is now made in terms of whether the 
energy level is {\em particle like}, or {\em phonon-like}.  We take 
the division between the bands to be at $E_R$, where $E_R$ is the 
energy above which all the excitations can be considered, to a good 
degree of approximation, to consist of a definite number of 
particles, unlike the phonon-like excitations, which have lower 
energies.

\Beginfigure
\infig{8.5cm}{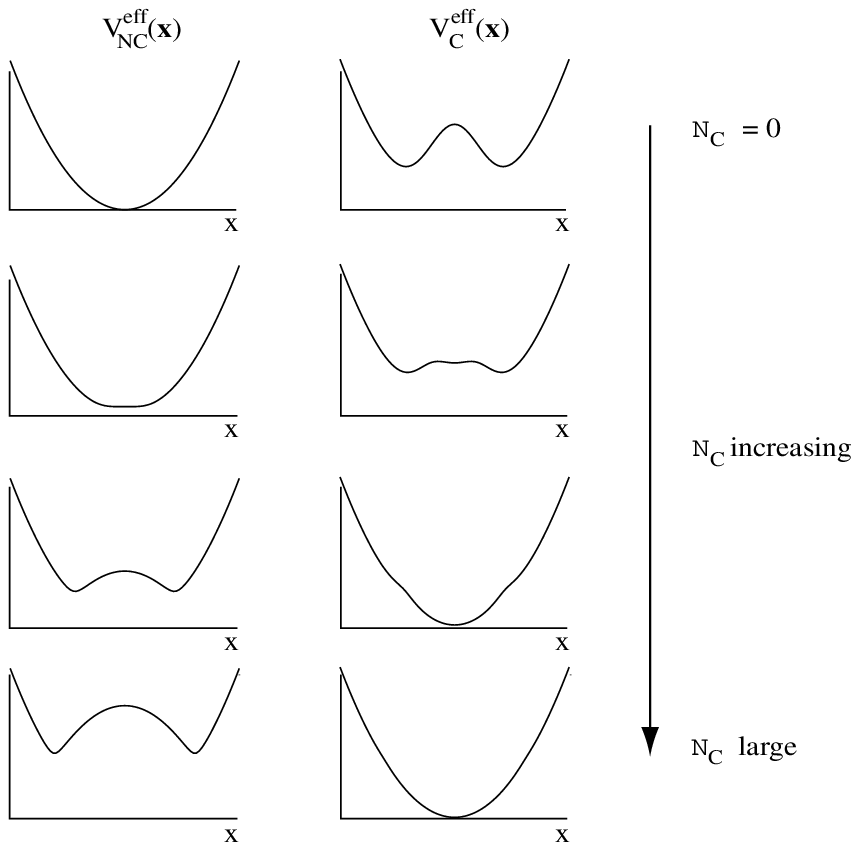} \Caption{Fig.~2: Representation of the 
non-condensate-band (left) and condensate band (right) effective potentials
appropriate to an increasing number of condensate atoms. Fore each particular 
sector, defined by the numbers $ N_{\rm NC}$ and $ N_{\rm C}$, the different 
effective potentials define the basis eigenstates in terms of which the master 
equation is defined. The process of growth is that of transition from one 
sector to the next, and does not involve a time dependence of the potential. }
\Endfigure

In order to take account of mean-field effects of the condensate-band atoms on 
the non-condensate-band atoms, and conversely, we use 
{\em effective trap potentials}, for the condensate and non-condensate bands
 which include these mean-field effects.  Thus, we  write
\begin{eqnarray}\Label{In6}
V^{\rm mf}_{\rm NC}({\bf x}) &=& 2u[\bar\rho({\bf x}) + \bar n({\bf x})],
\\ \Label{In601}
V^{\rm mf}_{\rm C}({\bf x}) &=& 2u\bar n({\bf x}),
\end{eqnarray}
where $ \bar\rho({\bf x})$ is the mean condensate band density, and 
$ \bar n({\bf x})$ is the mean density of the vapor in the non-condensate band.  
The values of these quantities depend on the particular physical situation, 
including the numbers of particles, $ N_{\rm NC}$ and $ N_{\rm C}$, in the 
respective bands. 

The {\em effective potentials} are then
\begin{eqnarray}\Label{In6a}
V^{\rm eff}_{\rm NC}({\bf x}) &=& V_T({\bf x}) + V^{\rm mf}_{\rm NC}({\bf x})
\\ \Label{In6b}
V^{\rm eff}_{\rm C}({\bf x}) &=& V_T({\bf x}) + V^{\rm mf}_{\rm C}({\bf x})
\end{eqnarray} 
As written, the effective potentials contain no time-dependence.  However, 
time-dependent situations can be dealt with in two ways.  

Firstly, consider condensate growth.  The picture put forward by us in 
\cite{BosGro,NewestBosGro} treats condensate growth as the slow change of the 
populations of the condensate and non-condensate bands by transitions of 
particles from one band to the other.  The transitions are between energy 
levels determined by the values of the densities $ \bar n({\bf x})$, 
$ \bar\rho({\bf x})$ appropriate to the actual numbers $ N_{\rm NC}$,
 $ N_{\rm C}$ of non-condensate-band and condensate-band atoms.  Thus the 
values of $ \bar n({\bf x})$ and $ \bar\rho({\bf x})$ (and hence also the 
mean-field ptentials) will change adiabatically as the condensate grows, and 
the energy levels needed to compute the transition rates will also change 
adiabatically.  The basic requirement is that the growth is sufficiently slow 
compared to the enrgy thermalization time within each band that there are 
always well-defined values of  $ \bar n({\bf x})$ and $ \bar\rho({\bf x})$.

The second way in which time-dependence can be treated is perturbatively.  If 
the condensate and vapor oscillate with small amplitude about mean values, we 
can show that this can be taken into account by using appropriate time 
dependent values of  $ \bar n({\bf x})$ and $ \bar\rho({\bf x})$.

These two effective potentials provide a basis within which 
we can develop the full quantum kinetic theory.
This is done by dividing the field operator into two parts
\begin{eqnarray}\Label{In7a}
\psi({\bf x}) = \phi({\bf x}) + \psi_{\rm NC}({\bf x})
\end{eqnarray}
which then correspond to the condensate band and non-condensate band
degrees of freedom.  We then write the Hamiltonian (\ref{In1}) as
\begin{eqnarray}\Label{In8}
H&=& H^{\rm eff} + \left(H-H^{\rm eff}\right),
\end{eqnarray}
in which $ H^{\rm eff} $ is the sum of the condensate-band and non-condensate 
band effective Hamiltonians; hence the non-condensate-band effective 
Hamiltonian 
is
\begin{eqnarray}\Label{In9}\nonumber
H^{\rm eff}_{\rm NC}&=& \int d^3{\bf x}\,\psi^\dagger({\bf x})
\left\{
-{\hbar^2\nabla^2\over 2m}+V_{\rm NC}^{\rm eff}({\bf x}) 
\right\}\psi({\bf x}),\\
\end{eqnarray}
and the condensate band effective Hamiltonian is
\begin{eqnarray}\Label{In10}\nonumber
H^{\rm eff}_{\rm C}&=& \int d^3{\bf x}\,\phi^\dagger({\bf x})
\left\{
-{\hbar^2\nabla^2\over 2m}
 + V_{\rm C}^{\rm eff}({\bf x})\right\}
 \nonumber \\
&& +{u\over 2} \int d^3{\bf x}\,\phi^\dagger({\bf x})\phi^\dagger({\bf x})
\phi({\bf x})\phi({\bf x}).
\end{eqnarray}
The quantities $ \bar\rho({\bf x})$ and $ \bar n({\bf x})$ are nominally the 
mean condensate-band and non-condensate-band densities, but the real criterion 
for their choice is to minimize the effective interaction term $ H-H^{\rm eff}$

\subsubsection{Basis states for the master equation}\Label{In-basis}
The non-condensate-band effective Hamiltonian is now written in terms of a 
wavelet basis, as follows.  We write
\begin{eqnarray}\Label{In11}
\psi_{\rm NC}({\bf x}) 
&=& \sum_{{\bf Q},{\bf r}}w_{\bf Q}({\bf x},{\bf r})A_{{\bf Q},{\bf r}}
\equiv \sum_{\bf Q}\psi_{\bf Q}({\bf x}).
\end{eqnarray}
Here $ A_{{\bf Q},{\bf r}} $ are destruction operators with the commutation 
relations
\begin{eqnarray}\Label{In12}
[A_{{\bf Q},{\bf r}},A^\dagger_{{\bf Q}',{\bf r}'}]=
\delta_{{\bf Q}{\bf Q}'}\delta_{{\bf r}{\bf r}'},
\end{eqnarray}
so that $ {\bf r}$ and $ {\bf Q}$ are discrete indices.  The wavelets
$ w_{\bf Q}({\bf x},{\bf r})$ are a complete orthonormal set, such that
\vskip 5pt\noindent
1. 	$ w_{\bf Q}({\bf x},{\bf r})\to 0$ when $ |{\bf r}-{\bf x}| $ is large.

\vskip 5pt\noindent
2. 	 $ w_{\bf Q}({\bf x},{\bf r})$ is a linear combination of trap 
eigenfunctions for the effective potential 
$ V_{T}({\bf x}) +V_{\rm NC}^{\rm mf}({\bf r})$ with energy eigenvalues in a 
range restricted to a neighborhood of $ \hbar^2{\bf Q}^2/2m$.

\vskip 5pt\noindent
3. 	 With each set of $ {\bf Q}$, $ {\bf r}$ we can associate a momentum 
$ \hbar{\bf K}({\bf Q},{\bf r})$ which is such that
\begin{eqnarray}\Label{In13}
{\hbar^2{\bf Q}^2\over 2m}= {\hbar^2{\bf K}^2\over 2m}
+ V^{\rm eff}_{\rm NC}({\bf r}).
\end{eqnarray}
For every direction of $ {\bf Q}$ we can also associate a direction of 
$ {\bf K}$; for example in a spherically symmetric harmonic trap $ {\bf Q}$
 and $ {\bf K}$ are parallel.  

In general a rule of association between the 
directions is not simple, but it can be specified implicitly.

\vskip 5pt\noindent
4. 	Using the function  $ {\bf K}({\bf Q},{\bf r})$ we can write the 
commutator
\begin{eqnarray}\Label{In14}\nonumber
[\psi_{\bf Q}({\bf x}),\psi^\dagger_{{\bf Q}'}({\bf x}')] &\approx &
\delta_{{\bf Q}{\bf Q}'}e^{i{\bf K}({\bf Q},{{\bf x}+{\bf x}'\over 2})
\cdot({\bf x}-{\bf x}')}g_{\bf Q}({\bf x},{\bf x}')\\
\end{eqnarray}
where the function $ g_{\bf Q}({\bf x},{\bf x}')$ is only significantly 
different from zero when $ {\bf x}\approx {\bf x}' $, and is normalized to one 
in both variables $ {\bf x}$ and ${\bf x}'$.

\vskip 5pt
This set of wavelets can be explicitly constructed in a number of cases, and we 
assume its existence in general.  It provides a resolution of the 
non-condensate field into discrete ``cells'' in phases space labeled by $ {\bf 
Q}$ and $ {\bf r}$.  The wavelet function $ w_{{\bf Q}, {\bf r}}({\bf x})$ 
corresponds to a particle approximately localized at $ {\bf r}$ in space and 
with momentum in a small range around $ \hbar{\bf K}({\bf Q},{\bf r})$.  Its 
total energy is thus  approximately $ E({\bf Q}) \equiv \hbar^2{\bf Q}^2/2m$, 
which is independent of $ {\bf r}$.  A particle in a wavelet state 
$ {\bf Q},{\bf r}$ will evolve under the influence of the Hamiltonian 
$ H^{\rm eff}_{\rm NC}$ into a linear combination of wavelets with the {\em 
same} $ {\bf Q}$, so that the evolution represents energy conserving and 
particle conserving flow in phase space.  We thus obtain a Liouvillian flow in 
this discrete phase space.

The evolution of the condensate band is described by (\ref{In10}), which 
differs from the full Hamiltonian only in that

\vskip 5pt\noindent
1. 	The mean-field potential of the vapor $ V^{\rm mf}_{\rm C}({\bf x})$ 
occurs.

\vskip 5pt\noindent
2. 	The operator $ \phi({\bf x}) $ is limited to the lower energy levels of the 
full Hamiltonian, and thus $ \phi({\bf x})$ has a non-local commutation 
relation, the degree of non-locality being of the same size as the average 
spatial width of the wavelets used to construct the non-condensate-band.
\vskip 5pt

Although we have described the wavelets by approximate properties, once the 
wavelets have been constructed, the description in terms of them is exact.  
However, in order to use this description some approximations are still 
necessary.  The simplest procedure---and the one we shall use here---is to use 
the number-conserving Bogoliubov model, but include the vapor mean-field 
potential $ V^{\rm mf}_{\rm C}({\bf x}) = 2 u \bar n({\bf x})$ as well as the 
trap potential.  This provides us with an approximate resolution of the 
non-condensate-band field operator in the form
\begin{eqnarray}\label{In15}
\phi({\bf x}) &=& A\left(\xi_{N_{\rm C}}({\bf x}) 
+ {1\over\sqrt{N_{\rm C}+1}}\chi({\bf x})\right)
\end{eqnarray}
in which

\vskip 5pt\noindent
1. 	$ A$ is a destruction operator for the total number of atoms in the 
condensate band;

\vskip 5pt\noindent
2. 	$N_{\rm C}\equiv A^\dagger A $ is the number of atoms in 
the condensate band;

\vskip 5pt\noindent
3. 	$\xi_{N_{\rm C}}({\bf x}) $ is the  condensate wavefunction.

\vskip 5pt\noindent
4. 	The quantity $ \chi({\bf x}) $ is a quasiparticle field operator, which 
has the expansion
\begin{eqnarray}\Label{In16}
 \chi({\bf x})= \sum_m\left\{p_m({\bf x}) b_m + q_m({\bf x})b^\dagger_m\right
\}.
\end{eqnarray}
\vskip 5pt

The $ p_m({\bf x}) $ and
$ q_m({\bf x}) $ are quasiparticle amplitudes for the quasiparticle creation 
and destruction operators $ b_m^\dagger, b_m$.  The condensate has energy 
$ \mu(N_{\rm C})$, and each quasiparticles carries an energy 
$ \hbar\epsilon_m(N_{\rm C})$. 

We assume that $ N_{\rm C}$ is almost equal to 
the number of particles $ N$ in the condensate level. This is a significantly 
weaker requirement than that which would be necessary to use the  usual 
Bogoliubov method for the whole system, which would require
the total number of particles in the system, $ N_{\rm C} + N_{\rm NC}$, to be 
close to the number of particles in the condensate.

At this degree of approximation, the description is rather like the 
conventional Hartree-Fock-Bogoliubov \cite{HFB} method.  However, there
is so far no thermalization assumed---this can only be done satisfactorily by 
introducing irreversibility, which we shall do using our master equation.

\subsubsection{The master equation}
The master equation has two main purposes

\vskip 5pt\noindent
1. 	 To take account of the interactions that are included in the term 
$ H-H^{\rm eff}$ in the equation (\ref{In9})
\vskip 5pt\noindent
2. 	 To introduce irreversibility, and thus thermalization.
\vskip 5pt
The technical methodology we use is very similar to that used in QKI. 
The equation of motion for the density operator is developed perturbatively by 
projecting onto the space of density operators in which the relative phases of 
non-condensate-band states with {\em different} $ {\bf Q}$ are eliminated.  We 
cannot eliminate the phases corresponding to the same $ {\bf Q}$ and different 
$ {\bf r}$, since the Liouvillian transport generated by $ H^{\rm eff}$ is 
off-diagonal in $ {\bf r}$, would itself be eliminated by such a procedure.  
But Liouvillian transport generates the streaming terms in the quantum 
Boltzmann equation, and must therefore be preserved if we are to obtain a full 
hydrodynamic description of the vapor.

As in QKI and QKIII the derivation of the master equation 
relies on a perturbation expansion in the interaction term, and on a random 
phase approximation in certain of the coefficients.  The master equation which 
results is natural, self-consistent and consistent with statistical mechanics. 
 
For example, a stationary solution for the density operator can be written
\begin{eqnarray}\Label{In17}
\rho_s &=& \exp\left(-{H_a^{\rm eff}-\mu N_{{\rm NC}}\over kT}\right)
\otimes    \exp\left(-{H_{\rm C}^{\rm eff}-\mu N_{{\rm C}}\over kT}\right).
\nonumber \\
\end{eqnarray}
Here 
\begin{eqnarray}\Label{In18}
H_a^{\rm eff}&=& 
\sum_{{\bf r},{\bf Q}}E({\bf Q})A^\dagger_{{\bf Q},{\bf r}}A_{{\bf Q},{\bf r}}.
\end{eqnarray}
and $ H_{\rm C}^{\rm eff} $ is given by (\ref{In10}).

There are thus no correlations between the condensate and non-condensate bands, 
and none between the different  $ {\bf Q},{\bf r}$ ``cells'' in the 
non-condensate-band.  This description is self-consistent if the highest energy 
states in the condensate-band (and all higher energy states) are essentially 
particle-like, since the density operator will then look much the same at the 
top of the condensate-band as at the bottom of the non-condensate band.
The master equation in the full form is written in Sect.\ref{ME terms}, and 
comprises terms which account for all physical processes as follows.
\vskip 3pt
\noindent
{\bf 1. Condensate-band term:} This is given by (\ref{bba201},\ref{bba20201}), 
and arises 
from the condensate band Hamiltonian $ H_{\rm C}^{\rm eff}$, (\ref{In10}).

\vskip 3pt
\noindent
{\bf 2. Non-condensate-band terms:}
These are given by  
(\ref{bba203a}--\ref{bba203e}), and are very like those in QKI.  They 
account for Liouvillian flow and collisions within the non-condensate-band.

\vskip 3pt
\noindent
{\bf 3.  Mean-field coupling term:}
One part of the interaction leads to a 
Hamiltonian coupling between the two bands, and this is a residual part of the 
Hamiltonian, which can be written
\begin{eqnarray}\Label{In19}
H_{\rm mf} &=& 2u\int d^3{\bf x} \Big\{\Big(
\sum_{\bf Q}\psi^\dagger_{\bf Q}({\bf x})\psi_{\bf Q}({\bf x})-\bar n({\bf x})
\Big)
\nonumber \\&&\times
\Big(\phi^\dagger({\bf x})\phi({\bf x}) -\bar \rho({\bf x})\Big)\Big\} .
\end{eqnarray}
Since this is only a residual after subtracting the mean-field terms, it is 
expected to be of the same order of magnitude as the irreversible terms.  
It is 
neglected entirely in writing the stationary density operator in the form 
(\ref{In17}). 
\vskip 3pt
\noindent
{\bf 4. Irreversible coupling terms:}
These are the irreversible terms which give rise to energy and particle 
transfer between the condensate-band and the non-condensate-band.  They are 
written in terms of some formal condensate-band operators, and the 
field operators $ \psi_{\bf Q}({\bf x})$ of the non-condensate-band.  In order 
to use the master equation, the eigenfunctions of $ H_{\rm C}^{\rm eff}$, the 
effective condensate-band Hamiltonian, must be found.  This is a question of 
choosing the appropriate approximation method.

\subsection{Practical methods}
By assuming that the non-condensate band can be maintained in a thermal 
stationary state we recover the methodology of QKIII.  The next obvious step is 
to take a time-dependent local equilibrium approximation for the 
non-condensate-band, and a number-conserving Bogoliubov approximation for the 
condensate-band.  We can then talk about the system as a thermalized vapor in 
the non-condensate-band, in interaction with a system composed of a condensate 
and quasiparticles in the condensate-band.  This is the most useful way of 
expressing our results.  We find that

\vskip 5pt\noindent
1. 	The condensate-band is described as the Bogoliubov system of quasiparticles 
and condensate, in which the mean-field potential induced by the vapor may be 
time-dependent.

\vskip 5pt\noindent
2. 	The vapor is described by a quantum Boltzmann equation, in which the mean 
field 
potential induced by the condensate band may be time-dependent.

\vskip 5pt\noindent
3. 	There are transfer terms between the condensate-band and vapor, which may 
be evaluated in various degrees of approximation.
\vskip 5pt

\subsubsection{Non-condensate-band equations of motion}
The equation of motion for the phase-space density $ f_{{\bf K}}({\bf x})$ of 
the non-condensate-band vapor can be written
\widetext
\begin{eqnarray}\Label{In20}
{\partial f_{\bf K}({\bf x})  \over \partial t} &=&
 \left({\hbar {\bf K}\cdot \nabla_{{\bf x}}\over m}
-{\nabla_{\bf x}V_{\rm NC}^{\rm eff}({\bf x},t)\cdot\nabla_{\bf K}
\over\hbar}
\right)f_{\bf K}({\bf x})
\nonumber\\
&& +\frac{2| u|^2}{h^2}
\int\limits_{R_{\rm NC}} d^3{\bf K}_2
\int\limits_{R_{\rm NC}} d^3{\bf K}_3
\int\limits_{R_{\rm NC}}d^3{\bf K}_4 
\delta ( {\bf K}+{\bf K}_2-{\bf K}_3-{\bf K}_4)
\delta (\omega +\omega_2-\omega_3-\omega_4)
\nonumber \\
&&\times \Big\{
f_{{\bf K}}( {\bf x}) f_{{\bf K}_2}( {\bf x} )
[ f_{{\bf K}_3}( {\bf x}) +1] [ f_{ {\bf K}_4}( {\bf x}) +1]
 - [f_{{\bf K}}( {\bf x}) +1] [ f_{{\bf K}_2}( {\bf x}) +1]
 f_{{\bf K}_3}({\bf x}) f_{{\bf K}_4}( {\bf x}) \Big\}
\nonumber \\
&& + \left.{\partial f_{\bf K}({\bf x})  \over \partial t}\right |_{1}
+\left.{\partial f_{\bf K}({\bf x})  \over \partial t}\right |_{2}
+\left.{\partial f_{\bf K}({\bf x})  \over \partial t}\right |_{3}.
\end{eqnarray}
\narrowtext
The final three terms represent transfers of energy and atoms to and from the 
condensate band in which respectively 1, 2 and 3 condensate-band field 
operators are involved in the matrix elements.

The transfer terms are given in full in (\ref{le6}), (\ref{701}) and 
(\ref{le9}), and are necessarily rather complicated because of the mixing of 
creation and destruction operators generated by the Bogoliubov method.  
However, 
the essence of the result is that there are simple rate processes involving 
non-condensate-band particle numbers, condensate band quasiparticle numbers, 
and the number of particles in the condensate.  All of the rates involve 
integrals like
\begin{eqnarray}\Label{In21}
&&
\int\!\! d^3 {\bf K}_2\!\!\int\!\! d^3 {\bf K}_3\!\!\int\!\! d^3 {\bf k}\,
\delta\left(\omega_{2{\bf K}4}({\bf x})-\omega\right)
\delta({\bf K}+{\bf K}_4-{\bf K}_2-{\bf k})
\nonumber \\  &&
\times\left[1 +f_{\bf K}({\bf x})\right]
      \left[1+f_{{\bf K}_4}({\bf x})\right]
     f_{{\bf K}_2}({\bf x}){\cal W}_I({\bf x},{\bf k}).
\end{eqnarray}
The quantity $ {\cal W}_I({\bf x},{\bf k}) $ is a Wigner function of a 
quasiparticle amplitude---we see therefore that there are energy and momentum 
conservation delta functions as in the quantum Boltzmann equation, and where a 
particle is in the condensate band we replace its phase space density by the 
Wigner function corresponding to the particular quasiparticle amplitude for the 
process being considered.  Thes amplitudes are closely related to the functions 
$ p_m({\bf x})$, and $ q_m({\bf x})$ in equation (\ref{In16}).  The effective 
potential $ V^{\rm eff}({\bf x})$ experienced by the vapor as a result of the 
existence of a condensate and its quasiparticles.
\begin{eqnarray}\Label{In22}
 V^{\rm eff}({\bf x})&=& 2u\bar\rho({\bf x})
\\ \nonumber 
&  = & 2uN|\xi_N({\bf x})|^2 
\\  && +
 \sum_m\left
\{\bar n_m |p_m({\bf x})|^2 +
(\bar n_m+1)|q_m({\bf x})|^2
\right\}
\end{eqnarray}
where the mean occupation number of the quasiparticle $ m$ is defined in terms 
of the operators of (\ref{In16}) and the condensate band density operator 
$ \rho_{\rm C}$:
\begin{eqnarray}\Label{In23}
\bar n_m &=& {\rm Tr}\left(b^\dagger_mb_m\rho_{\rm C}\right),
\end{eqnarray}
and is time-dependent if $ \rho_{\rm C} $ is time-dependent.

\subsubsection{Condensate-band equations of motion}
The methodology used does not demand that any further approximation be 
made to the condensate-band effective Hamiltonian as given in 
(\ref{In10}).  However the interactions with the non-condensate band 
are most naturally expressed in terms of quasiparticles, as explained 
in Sect.\ref{In-basis}, and this will only make sense when the 
condensate-band Hamiltonian is expressed in terms of the Bogoliubov 
theory.  To a first approximation it is reasonable to treat the 
condensate-band as a condensate plus a gas of non-interacting 
quasiparticles, and this is all we do explicitly in this paper.

There is no difficulty in principle in including the 
interactions between the quasiparticles which would arise in a full 
expansion of the condensate-band Hamiltonian using the quasiparticle 
states as a basis.  However a practical method of doing this 
would probably depend on the specific problem under consideration.

\section{Use of wavelets for for trapped atoms}
In QKI we introduced a wavelet expansion of the field operators.  The 
wavelets are a complete set of orthonormal one-particle wavefunctions 
which are only significantly different from zero in a phase space 
volume $h^3$ around a point $\hbar{\bf K},{\bf r}$ in phase space.  By 
this we mean that the wavelet function $v_{\bf K}({\bf x},{\bf r})$ is 
only significantly different from zero in a co-ordinate space volume
$\pi^3/ \Delta^3$ around $ {\bf r}$, and that their expression in terms of 
momentum eigenfunctions with eigenvalue $ {\bf p}$ is only significantly 
different from zero in a momentum space volume $ (2\hbar\Delta)^3 $ around
$ {\bf p}= \hbar{\bf K}$.  This expansion enables one to express the quantum 
states 
of the Bose gas in terms of the numbers of particles associated with the 
wavelet $ {\bf K},{\bf r}$; which we can loosely think of as the number of 
particles in a phase space ``cell'' of volume $ h^3$ at the point 
$ {\bf K},{\bf r}$.  This gives a fully quantum mechanical description of the 
physics, but also connects very directly the classical description of the 
particle as having position $ {\bf r} $, and momentum $ \hbar{\bf K}$.  Using 
the description, we can talk about scattering between different momentum values
(the Boltzmann collision term), and flow through phase space (the Boltzmann 
streaming term).  Because it is fully quantum mechanical, it is 
reasonably straightforward to use this phase space description for the 
non-condensate band $ R_{\rm NC}$ at the same time as we use a description in 
terms of energy eigenfunctions in the condensate band $ R_{\rm C}$.

\subsection{Formulation of wavelets}
\subsubsection{Wavelet properties}
The details of the construction of the wavelets are contained in 
Appendix \ref{AppA}.  The essential properties of the wavelets which we shall 
use here are as follows. We write the one-particle trapping Hamiltonian as 
\begin{eqnarray}\Label{in.wv1}
H_T^{(\rm one\ particle)}&=& {{\bf p}^2\over 2m} + V({\bf x}) 
\end{eqnarray}
we can choose a set of wavelet functions $ w_{\bf Q}({\bf x},{\bf r})$
with the following properties.

\vskip 5pt\noindent
1.
 The wavelets are a complete orthonormal set
\begin{eqnarray}\Label{in.wv2}
 \int d^3{\bf x}\,w^*_{\bf Q}({\bf x},{\bf r}) w_{\bf Q'}({\bf x},{\bf r'})
&=& \delta_{\bf QQ'}\delta_{\bf rr'},
\\
\sum_{\bf Q,r}w^*_{\bf Q}({\bf x},{\bf r}) w_{\bf Q}({\bf x'},{\bf r})
&=&\delta({\bf x}-{\bf x'}).
\end{eqnarray}

\vskip 5pt\noindent
2.  The wavelet function $  w_{\bf Q}({\bf x},{\bf r}) $ is only 
significantly different from zero when $ {\bf x}\approx {\bf r}$.

\vskip 5pt\noindent
3. The mean energy of the wavelet is independent of $ {\bf r}$, and is
\begin{eqnarray}\Label{in.wv3}
E({\bf Q}) &=& {\hbar^2{\bf Q}^2\over 2m}.
\end{eqnarray}
The mean momentum associated with a wavelet is a vector
$ \hbar{\bf K}({\bf Q},{\bf r})$ which is such that
\begin{eqnarray}\Label{in.wv4}
{\hbar^2{\bf Q}^2\over 2m} &=& 
{\hbar^2\big({\bf K}({\bf Q,r})\big)^2\over 2m} +V({\bf r})
\end{eqnarray}
The direction of $ {\bf Q}$ is determined by the condition that the Liouvillian 
flow in phase space does not change $ {\bf Q}$; this means that 
$ {\bf K}({\bf Q,r})$ can be inverted to give a function
$ {\bf Q}({\bf K,r})$, which can be regarded as a smooth function of 
continuous 
variable on a sufficiently large scale, and that this function satisfies the  
equation
\begin{eqnarray}\Label{in.wv5}
\left\{{\hbar{\bf K\cdot \nabla_{\bf r}}\over m}
-{\nabla V({\bf r})\cdot\nabla_{\bf K}\over \hbar}\right\}{\bf Q}({\bf K,r}) 
=0.
\end{eqnarray}
Clearly, this is compatible with (\ref{in.wv4}).  We can fix the function by 
requiring that
\begin{eqnarray}\Label{in.wv6}
V({\bf r})=0 &\quad\Longrightarrow\quad& {\bf Q}({\bf K,r})={\bf K}.
\end{eqnarray}
These requirements can be all satisfied simultaneously, as is demonstrated in 
Appendix \ref{AppA}.

\subsubsection{Field operator in terms of wavelets}
\vskip 5pt\noindent
1.
The field operator can be expressed as
\begin{eqnarray}\Label{in.wv7}
\psi({\bf x}) &=& \sum_{\bf Q,r}w_{\bf Q}({\bf x},{\bf r}) A_{\bf Q,r}
\\ \Label{in.wv701}
&=& \sum_{\bf Q}\psi_{\bf Q}({\bf x}),
\end{eqnarray}
and 
\begin{eqnarray}\Label{in.wv8}
[ A_{\bf Q,r}, A^\dagger_{\bf Q',r'}] &=& \delta_{\bf QQ'} \delta_{\bf rr'}.
\end{eqnarray}

\vskip 5pt\noindent
2.  For $ {\bf r,Q}$ sufficiently large that they may be regarded as 
being almost continuous variables, the commutator of the $ \psi_{\bf Q}({\bf 
x})$ 
functions has the asymptotic form
\begin{eqnarray}\Label{in.wv9}
&&[\psi_{\bf Q}({\bf x}),\psi^\dagger_{{\bf Q}'}({\bf x}')]
= \delta_{{\bf Q},{\bf Q}'}
\sum_{\bf r}w_{{\bf Q},{\bf r}}({\bf x})
            w^*_{{\bf Q},{\bf r}}({\bf x}')
\\ \Label{in.wv10}
&&\qquad\approx
\delta_{{\bf Q},{\bf Q}'}
e^{i{\bf K}\left({\bf Q},{{\bf x}+{\bf x}'\over 2}\right) 
\cdot({\bf x}-{\bf x}')}g_{\bf Q}({\bf x},{\bf x}').
\end{eqnarray}
where the function $ g_{\bf Q}({\bf x},{\bf x}') $ is only significantly 
different from 
zero when $ {\bf x}\approx{\bf x}'$, and is normalized to
\begin{eqnarray}\Label{in.wv11}
\int d^3{\bf x'}\,g_{\bf Q}({\bf x},{\bf x}')
=\int d^3{\bf x}\,g_{\bf Q}({\bf x},{\bf x}') = 1.
\end{eqnarray}
The region of validity of this asymptotic form coincides essentially with that 
of the WKB approximation for the wavefunctions.

\subsubsection{Expression of the many particle Hamiltonian}
\noindent
1.  The many particle {\em trapping Hamiltonian}---that is, excluding 
interactions (\ref{In3})---can be expressed as
\begin{eqnarray}
\Label{in.wv12}
H_T +H_{\rm kin} &=& \sum_{\bf r}\sum_{{\bf Q}}E({\bf Q})
       A^\dagger_{{\bf Q},{\bf r}}A_{{\bf Q},{\bf r}}
\nonumber \\ &&
+ \sum_{{\bf r}{\bf r}'}\sum_{{\bf Q}}
     {\cal M}({\bf Q},{\bf r},{\bf r}') 
      A^\dagger_{{\bf Q},{\bf r}}A_{{\bf Q},{\bf r}'}.
\end{eqnarray}

\vskip 5pt\noindent
2.
The object $  {\cal M}({\bf Q},{\bf r},{\bf r}')$ represents the 
quantized version of the Liouvillian motion, and defines a quantity
 $  M({\bf K},{\bf K}',{\bf r},{\bf r}')$ through
\begin{eqnarray}\Label{in.wv13}
 M\Big({\bf K}({\bf Q},{\bf r}),{\bf K}({\bf Q},{\bf r}'),{\bf r},{\bf r}'\Big)
&=& {\cal M}({\bf Q},{\bf r},{\bf r}'),
\end{eqnarray}
a definition which is only valid for $ ({\bf r},{\bf K}),({\bf r}',{\bf K}')$ 
on the same $ {\bf Q}$ 
surface.  When applied to a sufficiently smooth function $ F({\bf K},{\bf r})$, 
the $ M$ function generates the Liouvillian motion through
\begin{eqnarray}\Label{in.wv14}
&&\sum_{{\bf K',r'}} M({\bf K},{\bf K}',{\bf r},{\bf r}') F({\bf K}',{\bf r}')
\nonumber \\
&&\quad\approx i\hbar\left\{{\hbar{\bf K\cdot \nabla_{\bf r}}\over m}
-{\nabla V({\bf r})\cdot\nabla_{\bf K}\over \hbar}\right\} F({\bf K},{\bf r}).
\end{eqnarray}

\vskip 5pt\noindent
3.  The interaction part of the many-particle Ham\-il\-ton\-ian, 
$ H_{\rm I}$, (\ref{In3})
can be written   in the form
\begin{eqnarray}\Label{wv35}
H_{\rm I} &=&
 {1\over 2}\sum_{1234}{\cal U}_{{\bf Q}_1,{\bf Q}_2,{\bf Q}_3,{\bf Q}_4}
\end{eqnarray}
in which
\begin{eqnarray}\Label{wv36}
\nonumber
&& {\cal U}_{{\bf Q}_1,{\bf Q}_2,{\bf Q}_3,{\bf Q}_4}
\nonumber \\
&& =
\int d^3{\bf x} \int d^3{\bf x'}\,u({\bf x}-{\bf x}')
\psi^\dagger_{{\bf Q}_1}({\bf x)}\psi^\dagger_{{\bf Q}_2}({\bf x}')
\psi_{{\bf Q}_3}({\bf x}')\psi_{{\bf Q}_4}({\bf x})
\nonumber \\
\end{eqnarray}

\subsection{Definition of the condensate band and non condensate bands}
\Label{cond band}
The formulation in terms of wavelets so far has been appropriate for a 
situation in which no 
condensation occurs.  We now consider the situation in which there is 
significant condensation.  There are two aspects to be considered:

\vskip 5pt \noindent
1. 	We must  define the {\em condensate-band} and the 
{\em non-condensate-band} in terms of the $ {\bf Q}$-band formulation.  The 
condensate band, which we will call $ R_{\rm C}$, will consist of all $ {\bf 
Q}$ with energy less than a value $E_{\rm R}$, that is all $ {\bf Q}$ such that 
$ \hbar^2{\bf Q}^2/2m < E_{\rm R} $. 
The non-condensate band, which we will call $ R_{\rm NC}$, consists of the 
complement of $ R_{\rm C}$; that is, all states for which 
$ \hbar^2{\bf Q}^2/2m \ge E_{\rm R} $.
The {non-condensate-band} will be described in terms of wavelets, while the 
{\em condensate-band} will be described in terms of many-body eigenfunctions.

\vskip 5pt \noindent
2.  If there is significant condensation, there will be significant mean-field 
effects, which will give rise to a correction to the the trapping potential, as 
discussed in Sect.\ref{I.bands}.  For the non-condensate-band this means that 
the definition of the wavelets used will be changed, because these are defined 
in terms of a particular potential, as in (\ref{in.wv1}). 

\subsubsection{Commutation relations}
The procedure we will now follow is to resolve the field operators into 
$ {\bf Q}$-bands only within the non-condensate band, where we will find that 
the $ {\bf Q}$-bands are effectively thermalized, and are thus uncorrelated 
with each other.  In the condensate band, significant correlation 
effects may arise because the interactions with the condensate are not 
negligible
in this band---this leads to the use of  a single field operator the describe 
the behavior of the whole of $ R_{\rm C}$.

We now define the non-condensate band field operator $ \psi_{\rm NC}({\bf x})$, 
the 
field operator restricted to $ R_{\rm NC}$, based on the $ {\bf Q}$-band 
form for the field operator (\ref{in.wv7},\ref{in.wv701}), as 
\begin{eqnarray}\Label{bs1}
\psi_{\rm NC}({\bf x})&\equiv&
 \sum^{\rm NC}_{{\bf Q}}\sum_{\bf r}w_{\bf Q}({\bf x,r})A_{\bf Q,r}
\\ &\equiv&  \sum^{\rm NC}_{{\bf Q}}\psi_{\bf Q}({\bf x})
\end{eqnarray}
where the notation $ \sum^{\rm NC}_{{\bf Q}}$ means the summation over
$ {\bf Q}\in R_{\rm NC}$, with a corresponding notation for summations in $ R_{
\rm C}$.

The condensate band field operator will be called $ \phi({\bf x})$, and is 
defined by
\begin{eqnarray}\Label{defphi}
\phi({\bf x}) &\equiv& \psi({\bf x}) - \psi_{\rm NC}({\bf x})
\end{eqnarray}
The commutation relations for these operators are
\begin{eqnarray}\Label{bs101}
\left[\phi({\bf x}), \phi^\dagger({\bf x}')\right]
&=& \sum^{\rm C}_{{\bf Q}}
\sum_{\bf r}w_{{\bf Q},{\bf r}}({\bf x})
            w^*_{{\bf Q},{\bf r}}({\bf x}')
\nonumber \\
& \equiv &  g_{{\rm C}}({\bf x},{\bf x}').
\end{eqnarray}
The $ \phi$ and $ \psi_{\bf Q}$ operators commute with each other,
and the commutation relations of the $ \psi_{\bf Q}({\bf x})$ are as in 
(\ref{in.wv9},\ref{in.wv10})

\subsubsection{Separation of condensate and non-condensate parts of the full 
Hamiltonian}\Label{II.separation}
We can now write the full Hamiltonian in a form which separates the three 
components; namely, those which act within $ R_{\rm NC}$ only, those which act 
within $ R_{\rm C} $ only, and those which cause transfers of energy or 
population between $ R_{\rm C}$ and $ R_{\rm NC}$.  Thus we write the trapping 
plus kinetic parts of the Hamiltonian as the sum of three terms
\begin{eqnarray}\Label{H1}
H_{\rm kin}+H_T =  H_a + H_b  + H_{\rm C,1}
\end{eqnarray}
in which the part which gives the average energy of the non-condensate atoms
in each {\bf Q}-band  is
\begin{eqnarray}\Label{H2}
H_a &=& \sum_{\bf r}\sum^{\rm NC}_{{\bf Q}}{\hbar^2{\bf Q}^2\over 2m}
       A^\dagger_{{\bf Q},{\bf r}}A_{{\bf Q},{\bf r}}.
\end{eqnarray}
The part which represents Liouvillian transport of the non-condensate atoms
in each {\bf Q}-band  is 
\begin{eqnarray}\Label{H3}
H_b  &=& \sum_{{\bf r}{\bf r}'}\sum^{\rm NC}_{{\bf Q}}
     {\cal M}({\bf Q},{\bf r},{\bf r}')
      A^\dagger_{{\bf Q},{\bf r}}A_{{\bf Q},{\bf r}'}.
\end{eqnarray}
The part which represents the non-interaction part of the condensate-band 
Hamiltonian is
\begin{eqnarray}\Label{H6}
H_{\rm C,1} &=& \int d^3{\bf x}\,
\phi^\dagger({\bf x})
\left(-{\hbar^2\nabla^2\over 2m } +V_T({\bf x}) \right)
\phi({\bf x}).
\end{eqnarray}
The interaction part $ H_{\rm I}$ as defined in (\ref{In3}) can now be 
resolved into:

\vskip 5pt\noindent
1.
The part involving only $ \phi$ operators, which is the self interaction 
within $ R_{\rm C}$, which we call 
\begin{eqnarray}\Label{H02}
H_{\rm C,2}&\equiv &\int d^3{\bf x}\int d^3{\bf x}'{u({\bf x}-{\bf x}')\over 2}
\phi^\dagger({\bf x})\phi^\dagger({\bf x}')\phi({\bf x})\phi({\bf x}').
\nonumber \\
\end{eqnarray}

\vskip 5pt\noindent
2.
The part involving no $ \phi $ operators, which gives rise to 
scattering of the particles within $ R_{\rm NC}$, which we call 
\begin{eqnarray}\Label{HNC}
\nonumber
&&H_{\rm I,NC}\equiv
\\ \nonumber
&& \int d^3{\bf x}\int d^3{\bf x}'{u({\bf x}-{\bf x}')\over 2}
\psi_{\rm NC}^\dagger({\bf x})\psi_{\rm NC}^\dagger({\bf x}')\psi_{\rm NC}({\bf 
x})
\psi_{\rm NC}({\bf x}').
\nonumber\\
\end{eqnarray}

\vskip 5pt\noindent
3.
There are finally the terms involving operators from both bands, which cause 
transfer of energy and/or particles between $ R_{\rm C}$ and $ R_{\rm NC}$. We 
call the 
parts involving one $ \phi$ operator
\begin{eqnarray}\Label{HIC1}
\nonumber
&& H^{(1)}_{\rm I,C}\equiv 
\\ \nonumber
&&\,  \int d^3{\bf x}\int d^3{\bf x}'u({\bf x}-{\bf x}')
\psi_{\rm NC}^\dagger({\bf x})\psi_{\rm NC}^\dagger({\bf x}')\psi_{\rm NC}({
\bf 
x})
\phi({\bf x}') 
\\ \nonumber && \, + \int d^3{\bf x}\int d^3{\bf x}'u({\bf x}-{\bf x}')
\phi^\dagger({\bf x})\psi_{\rm NC}^\dagger({\bf x}')\psi_{\rm NC}({\bf x})
\psi_{\rm NC}({\bf x}').
\nonumber 
\\
\end{eqnarray}
The parts involving  two $ \phi $ operators are called
\begin{eqnarray}\Label{HIC2}
\nonumber
&& H^{(2)}_{\rm I,C}\equiv 
\\ \nonumber
&&\,  \int d^3{\bf x}\int d^3{\bf x}'{u({\bf x}-{\bf x}')}
\psi_{\rm NC}^\dagger({\bf x})\phi^\dagger({\bf x}')\psi_{\rm NC}({\bf x})
\phi({\bf x}')
\nonumber\\
&&+ \int d^3{\bf x}\int d^3{\bf x}'{u({\bf x}-{\bf x}')}
\phi^\dagger({\bf x})\psi_{\rm NC}^\dagger({\bf x}')\psi_{\rm NC}({\bf x})
\phi({\bf x}') 
\nonumber\\
&&+ \int d^3{\bf x}\int d^3{\bf x}'{u({\bf x}-{\bf x}')\over 2}
\psi_{\rm NC}^\dagger({\bf x})\psi_{\rm NC}^\dagger({\bf x}')\phi({\bf x})
\phi({\bf x}')
\nonumber\\
&&+ \int d^3{\bf x}\int d^3{\bf x}'{u({\bf x}-{\bf x}')\over 2}
\phi^\dagger({\bf x})\phi^\dagger({\bf x}')\psi_{\rm NC}({\bf x})
\psi_{\rm NC}({\bf x}'). \nonumber
\\
\end{eqnarray}
The parts involving  three $ \phi $ operators are called
\begin{eqnarray}\Label{HIC3}
\nonumber
&& H^{(3)}_{\rm I,C}\equiv 
\\ \nonumber
&&\,  \int d^3{\bf x}\int d^3{\bf x}'u({\bf x}-{\bf x}')
\phi^\dagger({\bf x})\phi^\dagger({\bf x}')\phi({\bf x})
\psi_{\rm NC}({\bf x}') 
\\ \nonumber &&\, + \int d^3{\bf x}\int d^3{\bf x}'u({\bf x}-{\bf x}')
\psi_{\rm NC}^\dagger({\bf x})\phi^\dagger({\bf x}')\phi({\bf x})
\phi({\bf x}').
\nonumber 
\\
\end{eqnarray}
Our notation is not entirely systematic, since $ H_{\rm I, NC}$ involves
only the non-condensate-band operators, whereas $ H_{\rm I,C}$ involves 
operators from both the condensate-band and the non-condensate-band. 

\subsubsection{Mean field corrections}\Label{MeanFieldCorr}
In practice the thermalized part of the condensate-vapor system consists of 
nearly all of the levels above the condensate level, and it is advantageous to 
include this fact explicitly as early as possible in the analysis. 
Thus, we would like to put the division between the condensate and 
non-condensate bands at as low an energy as possible.  
The main difference between the condensate band and the non-condensate band is 
the nature of the excitation spectrum.  It is relatively simple to include 
corrections to the energy levels of the non-condensate band provided these 
energy levels are particle-like---that is, they are states with a definite 
energy and a definite number of particles.  In fact, computations have shown 
only a small proportion of the excitation spectrum is not-particle like, and 
this is at quite low energy \cite{Stringari excitations}.  However, the 
energies of lower energy particle-like excitations are affected by the 
presence of the condensate and of each other.  We therefore must consider mean 
field effects. There are  three kinds of mean field effect which can be 
explicitly included:

\vskip 5pt\noindent
1. The average effect of the condensate band on  the non-condensate 
band;

\vskip 5pt\noindent
2.  The average effect of the non-condensate band on itself;

\vskip 5pt\noindent
3.  The average effect of the the non-condensate band on the condensate 
band.

\vskip 5pt\noindent
Our aim will be to estimate these average effects explicitly, and include them 
in the two Hamiltonians used to describe the two bands.  The sum of these two 
terms is then subtracted from the interaction part of the Hamiltonian, which is 
preserved as a correction term.  The description is thus in principle exact, 
since the total Hamiltonian is not changed---only the basis for the 
perturbation theory is changed.

If we suppose that $ \bar\rho({\bf x})$ is a c-number estimate of the 
condensate 
band density, and $ \bar n({\bf x})$ is a c-number estimate of the 
non-condensate 
band density, then the mean field effects 1.\ and 2.\ give an equivalent 
potential acting on the non-condensate
\begin{eqnarray}\Label{mf.01}
V_{\rm NC}^{\rm mf}({\bf x}) = 2u\Big(\bar\rho({\bf x})+  \bar n({\bf x})\Big)
\end{eqnarray}
while the effect 3. is given by an equivalent potential acting on the 
condensate 
\begin{eqnarray}\Label{mf.02}
V_{\rm C}^{\rm mf}({\bf x}) = 2 u  \bar n({\bf x})
\end{eqnarray}

The mean field effect is included by adding and subtracting a term 
\begin{eqnarray}\Label{mf.2}
&&\int d^3{\bf x}\left\{V_{\rm NC}^{\rm mf}({\bf x})
\psi^\dagger_{\rm NC}({\bf x})\psi_{\rm NC}({\bf x})
+ V_{\rm C}^{\rm mf}({\bf x})\phi^\dagger({\bf x})\phi({\bf x})\right\}.
\nonumber \\
\end{eqnarray}
Effectively, this changes the potentials used in defining the two bands.
For the non-condensate band, the change is
\begin{eqnarray}\Label{mf.3}
 V_T({\bf x})\to 
 V_{\rm NC}^{\rm eff}({\bf x})= V_T({\bf x})+V_{\rm NC}^{\rm mf}({\bf x}),
\end{eqnarray}
with a consequent redefinition of wavelets.  Thus the non-condensate band 
Hamiltonian
retains the {\em form} as in (\ref{H2},\ref{H3}), but the wavelets are now 
defined in terms of the {\em effective potential} 
$  V_{\rm NC}^{\rm eff}({\bf x}) $.

For the condensate the change is
\begin{eqnarray}\Label{mf.301}
 V_T({\bf x})&\to &
 V_{\rm C}^{\rm eff}({\bf x})= V_T({\bf x})+V_{\rm C}^{\rm mf}({\bf x}).
\end{eqnarray}
which now transforms the  $ H_{\rm C}$ to $ H^{\rm eff}_{\rm C}$, defined by
\begin{eqnarray}\Label{mf.302}
H_{\rm C}^{\rm eff}&= & 
 \int d^3{\bf x}\,
\phi^\dagger({\bf x})
\left(-{\hbar^2\nabla^2\over 2m } +V^{\rm eff}_{\rm C}({\bf x}) \right)
\phi({\bf x}) +H_{\rm C,2}.
\nonumber\\
\end{eqnarray}
The extra terms are then subtracted off the interaction part of the 
Hamiltonian, giving
\begin{eqnarray}\Label{mf.4}
 H^{(2)}_{\rm I,C} &\to  &  H^{(2,\rm eff)}_{\rm I,C}
\nonumber \\
&=&
 H^{(2)}_{\rm I,C}  - \int d^3{\bf x}\,V^{\rm mf}_{\rm NC}({\bf x})
\psi^\dagger_{\rm NC}({\bf x})\psi_{\rm NC}({\bf x})
\nonumber \\
&&-\int d^3{\bf x}\,V^{\rm mf}_{\rm C}({\bf x})\phi({\bf x})
\phi^\dagger({\bf x}).
\end{eqnarray}
We use the term $ H_{\rm I,C}^{\rm eff}$ for the sum of all of these 
interaction terms;
\begin{eqnarray}\Label{HIC}
H_{\rm I,C}^{\rm eff}\equiv H^{(1)}_{\rm I,C}+ 
H^{(2,\rm eff)}_{\rm I,C}+H^{(3)}_{\rm I,C}.
\end{eqnarray}
The description is still exact, since the mean-field terms have been added and 
subtracted, but an estimate of the mean effect of the interaction has been 
explicitly included in the basis description of the condensate and  
non-condensate.
The methodology does not depend on the particular choices of 
(\ref{mf.01},\ref{mf.02}). It is possible, for example 

\vskip 5pt\noindent
1. To choose the mean field potentials to be fixed quantities;

\vskip 5pt\noindent
2. To choose them to depend on the numbers of particles in the 
condensate and non-condensate bands;

\vskip 5pt\noindent
3. To include an explicit time dependence.  However, if this is done, 
the derivation and form of the master equation in Sect.\ref{III} would be 
altered.
\vskip 5pt

The choice of the estimates for $ \bar n({\bf x})$ and $ \bar\rho({\bf x})$
is not made at this stage, but will be done as part of the derivation of the 
master equation, since their choice must be such as to optimize 
the basis in which the master equation is described.

\section{Derivation of the master equation}\label{III}
Using the separation into condensate band and non-condensate band operators we 
can now proceed to develop a master equation in a a way that combines  
the methodologies of QKI and QKIII.
\subsection{Projectors in the {\bf Q}-bands}
We can now introduce the definitions of the projectors which we shall need, in 
much the same way as in QKI.
The states of the system will be written in the form
\begin{eqnarray}\Label{stat1}
|\Psi_{\rm C},{\bf n}\rangle
\end{eqnarray}
in which $ \Psi_{\rm C}$ represents the condensate degrees of freedom, 
and $ {\bf n}$ 
is a vector of integers $ n_{{\bf Q},{\bf r}}$ representing the numbers of 
atoms in the $ {\bf Q}$-cell $ {\bf Q},{\bf r} $.
We will want to consider a projector definition of the form
\begin{eqnarray}\Label{tp11}
p_{\bf N}|\Psi_{\rm C},{\bf n}\rangle& = &|\Psi_{\rm C},{\bf n}\rangle 
\qquad\mbox{if }
\sum_{\bf r}{\bf n}({\bf Q}',{\bf r})
   	 = N({\bf Q}),
\nonumber \\
& = & 0  \,\,\qquad \qquad\mbox{ otherwise.}
\end{eqnarray}
This projector projects only in $ R_{\rm NC}$, leaving the condensate 
state description $ \Psi_{\rm C}$ completely unaffected.

The projectors on the density operator are defined as in QKI
\begin{eqnarray}\Label{bs2001}
{\cal P}_{\bf N}\rho &\equiv & p_{\bf N}\rho\, p_{\bf N}
\\ \Label{bs2002}
&\equiv & v_{\bf N}.
\end{eqnarray}
The complementary projector $ {\cal Q}$ is defined by
\begin{eqnarray}\Label{bs200101}
{\cal Q} = 1 -\sum_{\bf N}{\cal P}_{\bf N}
\end{eqnarray}
where, by the definition (\ref{tp11}) of $ {\cal P}_{\bf N} $, the range of the 
index $ {\bf N}$ includes configurations entirely within $ R_{\rm NC}$.

Thus the terms $ v_{\bf N} $ will be the parts of the density operator for 
which we can derive a master equation, while the complementary part
\begin{eqnarray}\Label{bs200102}
w \equiv {\cal Q}\rho
\end{eqnarray}
is the part which is neglected.
 
The quantities $ v_{\bf N}$ are thus density operators in which $ {\bf N}$
represents the $ {\bf Q}$-band configuration of non-condensate-band particles, 
and in which no restriction of any kind is placed on the configuration of the 
particles in the condensate band.

\subsection{Formal derivation of the master equation}
The formal derivation of the master equation follows much the same methodology 
as in QKI and QKIII.  In the present situation, for the purposes of developing 
the master equation we can divide the operators into the various parts as in 
Sec.\ref{II.separation}, but also taking account of the fact that we use the 
effective Hamiltonians which include estimates of the mean field effects:
\begin{mathletters}
\begin{eqnarray}\Label{H7}
H_{\rm C}^{\rm eff}       = H_{\rm C,1}^{\rm eff}+H_{\rm C,2}
&&\qquad\mbox{[see (\ref{mf.302})]},
\\
H_{\rm I}^{\rm eff}      = H^{\rm eff}_{\rm I,NC}+ H^{\rm eff}_{\rm I,C}
&&\qquad\mbox{[see (\ref{mf.4},\ref{HIC})]},
\\
H_{\rm NC}^{\rm eff}  = H_a^{\rm eff}+H_b^{\rm eff}.
&&
\end{eqnarray}
\end{mathletters}
In the last of these, we understand $ H_a^{\rm eff} $ and $H_b^{\rm eff}$ to be 
constructed in the same way as (\ref{H2},\ref{H3}), but using the effective 
potential $ V^{\rm eff}_{\rm NC}({\bf x})$ innstead of $ V({\bf x})$ in 
(\ref{in.wv1}) as a basis for counstructing the wavelets.

There are the relations
\begin{mathletters}
\begin{eqnarray}\Label{H701a}
 [H^{\rm eff}_a,{\cal P}_{\bf N}\rho] 
&=& 0
\\ \Label{H701b}
{\cal P}_{\bf N}H^{\rm eff}_b\rho &=& H_b^{\rm eff}{\cal P}_{\bf N}\rho
\\ \Label{H701c}
{\cal P}_{\bf N}H^{\rm eff}_{\rm C}\rho &=& H_{\rm C}^{\rm eff}{\cal P}_{\bf N}
\rho
\end{eqnarray}
\end{mathletters}
which follow directly from the construction of the operators and projectors.

The equation of motion is 
\begin{eqnarray}\Label{H8}
\dot \rho &=& -{i\over \hbar}
[H_{a}^{\rm eff} + H_b^{\rm eff} +H^{\rm eff}_I +H^{\rm eff}_{\rm C}, \rho]
\\ &\equiv &  ({\cal L}_{a}+{\cal L}_{b}+{\cal L}_I +{\cal L}_{\rm C}) \rho .
\end{eqnarray}
Notice that, for compactness, we do not explicitly write $ {\cal L}^{\rm eff}$ 
for the Liouvillians.

From this form it is straightforward to write the master equation in the same 
form as in QKI.  As in QKI, we use the Laplace transform notation for any 
function $ f(t)$ 
\begin{eqnarray}\Label{Laplace}
\tilde f(s) &=& \int_0^\infty e^{-st}f(t)\,dt.
\end{eqnarray}
Using this notation and the relations (\ref{H701a}--\ref{H701c}), the master 
equation takes the form
\begin{eqnarray}\Label{H9.01}
&&\quad  s \tilde v_{\bf N}(s) -  v_{\bf N}(0) =
{\cal L}_{b}\tilde v_{\bf N}(s) 
+ {\cal P}_{\bf N}{\cal L}_I\sum_{\bf M}\tilde v_{\bf M}(s)
\nonumber \\ &&\qquad
+{\cal P}_{\bf N}{\cal L}_I[s- {\cal L}_{a}-{\cal L}_{b} -{\cal L}_{\rm C}
-{\cal Q}{\cal L}_I]^{-1}{\cal Q}{\cal L}_I\sum_{\bf M}\tilde v_{\bf M}(s)
\nonumber \\
\end{eqnarray}
In this form the master equation is basically exact.  We shall make the 
approximation that the kernel of the second part, the $ [\quad]^{-1}$ term, can
be approximated by keeping only the terms which describe the effective 
Hamiltonians within $ R_{\rm C}$ or $ R_{\rm NC}$, namely the terms 
$ {\cal L}_{a} $ and 
$ {\cal L}_{\rm C}$. 
Thus the master equation we shall use is written
\begin{eqnarray}\Label{H9}
&&\quad  s \tilde v_{\bf N}(s) -  v_{\bf N}(0) =
{\cal L}_{b}\tilde v_{\bf N}(s) 
+ {\cal P}_{\bf N}{\cal L}_I\sum_{\bf M}\tilde v_{\bf M}(s)
\nonumber \\ &&\qquad
+{\cal P}_{\bf N}{\cal L}_I[s- {\cal L}_{a}-{\cal L}_{\rm C}]^{-1}
{\cal Q}{\cal L}_I\sum_{\bf M}\tilde v_{\bf M}(s)
\end{eqnarray}
 This is essentially the same approximation as in QKI, apart 
from the inclusion of the condensate band self-interaction term in 
$ {\cal L}_{\rm C}$. However it should be noted that $ R_{\rm C}$ is a rather 
wide 
band 
compared to the the $ {\bf K}= 0$ band of QKI, and contains significant 
dynamics, which we do not assume to be thermalized---rather, we expect these 
levels to become thermalized as a consequence of the equations of motion.
\subsection{Reversible terms}

\subsubsection{Streaming Term}
This arises from the term $ {\cal L}_{b}v_{\bf N}(t)$, and takes an 
essentially 
classical form, based on of (\ref{in.wv12}--\ref{in.wv14}), in the degree of 
approximation we shall later use.

\subsubsection{Mean-field and forward scattering terms}
These arise from the term 
${\cal P}_{\bf N}{\cal L}_I\sum\limits_{\bf M}\tilde v_{\bf M}( s) $.  
Using the quantity $ {\cal U}_{{\bf Q}_1,{\bf Q}_2,{\bf Q}_3,{\bf Q}_4} $ as 
defined in (\ref{wv36}), we find that these terms are equivalent to a 
Hamiltonian term of the form
\begin{eqnarray}\Label{H11}
&&H_{\rm forward}\equiv
{1\over 2} \sum\limits_{{\bf Q}_1 \ne {\bf Q}_2}
\bigg\{{\cal U}_{{\bf Q}_1,{\bf Q}_2,{\bf Q}_1,{\bf Q}_2}
+{\cal U}_{{\bf Q}_1,{\bf Q}_2,{\bf Q}_2,{\bf Q}_1} \bigg\}
\nonumber \\
&&\quad + \sum\limits_{{\bf Q}}\Bigg\{
{1\over 2}{\cal U}_{{\bf Q},{\bf Q},{\bf Q},{\bf Q}}
\nonumber \\
&&\quad
+\int d^3{\bf x} \int d^3{\bf x'}\,{u({\bf x}-{\bf x}')}
\psi^\dagger_{{\bf Q}}({\bf x)}\phi^\dagger({\bf x}')
\phi({\bf x}')\psi_{{\bf Q}}({\bf x})
\nonumber \\
&&\quad +\int d^3{\bf x} \int d^3{\bf x'}\,{u({\bf x}-{\bf x}')}
\psi^\dagger_{{\bf Q}}({\bf x)}\phi^\dagger({\bf x}')
\phi({\bf x})\psi_{{\bf Q}}({\bf x}')
\nonumber\\
&&\quad 
- \int d^3{\bf x}V^{\rm mf}_{\rm NC}({\bf x})
\psi^\dagger_{{\bf Q}}({\bf x})\psi_{{\bf Q}}({\bf x})
\Bigg\}
\nonumber\\
&&\quad
- \int d^3{\bf x}V^{\rm mf}_{\rm C}({\bf x})
\phi^\dagger({\bf x})\phi({\bf x})
.
\end{eqnarray}
In  (\ref{H11}) one can see clearly the ``Hartree'' term in the third line, 
and 
the ``Fock'' term in the fourth line.  However in the remainder of this paper 
we shall use  the delta function form of the interaction potential
$ u({\bf x}-{\bf x}')\to u\delta({\bf x}-{\bf x}')$, and the difference between 
these terms and the second and third lines of (\ref{HIC2}) disappears.  None 
of 
our results will depend heavily on the validity of this approximation, but the 
formulae do become considerably more compact when it is made.

%

We thus arrive at the final form of the forward scattering term
\begin{eqnarray}\Label{mf.a3}
&& H_{\rm forward} \to H_{\rm forward}^{\rm eff} 
\nonumber \\
&& =
 \sum\limits_{{\bf Q}_1 \ne {\bf Q}_2}
{\cal U}_{{\bf Q}_1,{\bf Q}_2,{\bf Q}_1,{\bf Q}_2}
 + \sum\limits_{{\bf Q}}\Bigg\{
{1\over 2}{\cal U}_{{\bf Q},{\bf Q},{\bf Q},{\bf Q}}
\nonumber \\
&&\quad
+2u\int d^3{\bf x} 
\psi^\dagger_{{\bf Q}}({\bf x)}\psi_{{\bf Q}}({\bf x})
\left\{
\phi^\dagger({\bf x})\phi({\bf x})- \bar\rho({\bf x})-\bar n({\bf x})
\right\}
\Bigg\}
\nonumber\\
&&\quad
- 2 u\int d^3{\bf x}\,\bar n({\bf x})\phi^\dagger({\bf x})\phi({\bf x})
\end{eqnarray}

\subsection{Collisional terms}
The remaining parts of the master equation (\ref{H9.01}) lead to genuinely 
irreversible collision terms, both within the non-condensate band, and between 
the two bands.
\subsubsection{Collisions within the non-condensate band}
The interactions which involve only particles in 
$ R_{\rm NC}$ can cause collisions between the particles in the 
noncondensate band. Thus we consider the parts of the second line 
of (\ref{H9}) which arise from the part 
\begin{eqnarray}\Label{H20}
{\cal P}_{\bf N}{\cal L}_{\rm I,NC}[s- {\cal L}_{a} -{\cal L}_{\rm C}]^{-1}
{\cal Q}{\cal L}_{\rm I,NC}\sum_{\bf M}v_{\bf M}(s).
\end{eqnarray}
The result of this process can be seen by considering a particular term 
in the expansion of (\ref{H20}).  The inversion of the Laplace transform 
would give a term like
\begin{eqnarray}\Label{bba4} 
&&{\cal U}(1234)\int_0^\infty d\tau\,
\exp\{( {\cal L}_{a}+{\cal L}_{\rm C})\tau\}{\cal U}(4'3'2'1')v_{\bf N}(t-
\tau).
\nonumber \\&&
\end{eqnarray}
The only surviving term after 
projection will have 
\begin{eqnarray}
1=1',2=2' &\mbox{ or }&1=2', 2=1' \nonumber\\ &\mbox{and}& \nonumber\\
3=3',4=4' &\mbox{ or }&3=4', 4=3' \nonumber
\end{eqnarray}
and we can write this as
\begin{eqnarray}\Label{bba6} 
&&
{\cal U}(1234)\int_0^\infty d\tau\,e^{-i\Delta\omega_{1234}\tau}{\cal U}(4321)
\nonumber \\
&&\quad\times
e^{-{i\over\hbar}( H_{a}+H_{\rm C}^{\rm eff})\tau}
v_{\bf N}(t-\tau)
e^{{i\over\hbar}( H_{{a}}+H_{\rm C}^{\rm eff})\tau},
\end{eqnarray}
where, as in QKI,
\begin{eqnarray}\Label{bba601}
\Delta\omega_{1234} \equiv \omega_4+\omega_3-\omega_2-\omega_1 ,
\end{eqnarray}
but where we  use the notations
\begin{eqnarray}\Label{bba60101}
\hbar\omega_{\bf Q}& =& {\hbar^2{\bf Q}^2\over 2m} ,
\\
\omega_i &\equiv & \omega_{{\bf Q}_i} .
\end{eqnarray}
We now make the usual approximation that, in the time interval over which the 
$ \tau$ integration is significant, we may ignore the irreversible part of the 
time evolution of $ v_{\bf N}$, so that the exponential terms simply set 
$ t -\tau \to  t $, and we get simply
\begin{eqnarray}\Label{bba7}
{\cal U}(1234)\int_0^\infty d\tau\,e^{-i\Delta\omega_{1234}\tau}\,{\cal 
U}(4321)
v_{\bf N}(t).
\end{eqnarray}
which leads to the terms in the QKME which describe the scattering of the modes 
in $ R_{\rm NC}$ in exactly the same form as in (QKI:75d,e).

\subsubsection{Interaction between $ R_{\rm C}$  and $ R_{\rm NC}$}
The treatment of these terms depends on writing the interaction Hamiltonian 
terms in a form which uses wavelet functions for the non-condensate-band 
operators, and energy eigenfunctions for the condensate band trap operators as 
follows:
\begin{eqnarray}\Label{CNC1}
H^{(1)}_{\rm I,C}&=&
\sum_{234}\int d^3{\bf x}\,Z_{432}({\bf x})\sum_mX_m^\dagger({\bf x})
+ {\rm h.c.}
\\ \Label{CNC2}
H^{(2,\rm eff)}_{\rm I,C} &\approx&
2\sum_{42}\int d^3{\bf x}\,Z_{42}({\bf x})\sum_rU^\dagger_r({\bf x})
+{\rm h.c.}
\\ \Label{CNC3}
H^{(3)}_{\rm I,C} &=&
 \sum_{4}\int d^3{\bf x}\,Z_{4}({\bf x})
\sum_rV^\dagger_r({\bf x})
+{\rm h.c.}
\end{eqnarray}
The non-condensate band is represented by the $ Z$ operators; an abbreviated 
notation for
\begin{eqnarray}\Label{CNC4}
Z_{432}({\bf x}) 
&=& u \psi_4 ({\bf x})\psi_3 ({\bf x})\psi^\dagger_2 ({\bf x}),
\\ \Label{CNC5}
Z_{42}({\bf x})&=&  u \psi_4 ({\bf x})\psi^\dagger_2 ({\bf x}),
\\ \Label{CNC6}
Z_{4}({\bf x})&=&  u\psi_4 ({\bf x}).
\end{eqnarray}
The condensate band operators are defined by
\begin{eqnarray}\Label{CNC7}
\phi({\bf x}) &=& \sum_m X_m({\bf x})
\\ \Label{CNC8}
\phi({\bf x})\phi^\dagger({\bf x})& =& \sum_rU_r({\bf x}),
\\ \Label{CNC9}
\phi^\dagger({\bf x})\phi({\bf x})\phi({\bf x})
&=&  \sum_rV_r({\bf x}).
\end{eqnarray}
The $ X,U,V$ operators are, as in QKIII (where they are discussed more 
extensively),  eigenoperators of the condensate-band effective 
Hamiltonian, with eigenvalues defined by
\begin{eqnarray}\Label{CNC10}
[H_{\rm C}^{\rm eff}, X_m({\bf x})] = - \hbar \epsilon_m X_m({\bf x}),
\\ \Label{CNC11}
[H_{\rm C}^{\rm eff}, U_r({\bf x})] = - \hbar \beta_r U_r({\bf x}),
\\ \Label{CNC12}
[H_{\rm C}^{\rm eff}, V_r({\bf x})] = - \hbar \beta'_r V_r({\bf x}).
\end{eqnarray}
In the Bogoliubov approximation, these operators have relatively simple 
expressions in terms of quasiparticle creation and destruction operators---the 
precise relationship is given in Sec.IID of QKIII; especially
(94--99) which gives the basic definition, and Sec.IID3 which gives the 
expression of the more complicated $ U_r({\bf x})$ and $ V_r({\bf x})$ 
operators.

We omit terms involving  $ \phi({\bf x})\phi({\bf x})$, or its Hermitian
conjugate, since these are always very non-resonant.  We use the rotaing wave, 
or random phase approximation as in QKIII, and derive a master equation which 
is a blend of the formalism of QKI and that of QKIII.

\subsection{Master equation terms}\Label{ME terms}
Putting these all together we get the contribution to the master equation 
which describes coupling between $ R_{\rm C}$ and $ R_{\rm NC}$ so that 
the full quantum kinetic master equation now takes the form
\begin{eqnarray}\Label{bba201}
\dot v_{\bf N}(t) &=& 
  \dot v_{\bf N}(t)|_{\rm C}
+ \dot v_{\bf N}(t)|_{{\rm NC}}
+ \dot v_{\bf N}(t)|_{\rm MF}
+ \dot v_{\bf N}(t)|_{\rm coupling}
\nonumber \\
\end{eqnarray}
in which the various terms are defined as follows.
\subsubsection{Condensate band term}
This is 
\begin{eqnarray}\Label{bba202}
\dot v_{\bf N}(t)|_{\rm C} =
\left[H_{\rm C}^{\rm eff},v_{\bf N}(t) \right]
\end{eqnarray}
where, as in (\ref{mf.302}),
\widetext 
\begin{eqnarray} \Label{bba20201}
H_{\rm C}^{\rm eff} &=& \int d^3{\bf x}\,\,
\phi^\dagger({\bf x})
\left(-{\hbar^2\nabla^2\over 2m } +V^{\rm eff}_{\rm NC}({\bf x}) \right)
\phi({\bf x})+
{u\over 2}\int d^3{\bf x}\,\phi ^{\dagger }({\bf x})
\phi ^{\dagger }( {\bf x})\phi ( {\bf x}) \phi ( {\bf x}).
\end{eqnarray}
\subsubsection{Non-condensate band term}
\begin{mathletters}
\begin{eqnarray}\Label{bba203a} 
 \dot v_{{\bf N}}( t)|_{{\rm NC}} &=& 
-{ i\over\hbar}\left[H_b,v _{{\bf N}}(t) \right] 
\\ \Label{bba203c}
&&  - {i\over\hbar} \left[H_{\rm F,NC},v_{{\bf N}}(t)\right] 
\\  \Label{bba203d}
&& {- {i\over\hbar ^2}\sum\limits_{{\bf e}}}{\rm P}
{1\over\Delta\omega({\bf e})} \left[{\cal U}({\bf e})
{\cal U}^{\dagger }({\bf e}),
v_{\bf N}(t) \right] 
\\  \Label{bba203e}
&&{ +{\pi \over\hbar^2}{\sum\limits_{{\bf e}}}
\delta(\Delta\omega({\bf e}))}\{2{\cal U}({\bf e})v_{{\bf N-e}}(t)
{\cal U}^{\dagger}({\bf e}) 
-{\cal U}^{\dagger}({\bf e}){\cal U}({\bf e})v_{{\bf N}}(t)
-v_{{\bf N}}(t){\cal U}^{\dagger}({\bf e}) {\cal U}({\bf e})\}.
\end{eqnarray}
\end{mathletters}
Here we define the quantity 
\begin{eqnarray}\Label{bba230f}
H_{\rm F,NC} &=&   {1\over 2}\sum\limits_{{\bf Q}}
{\cal U}_{{\bf Q},{\bf Q},{\bf Q},{\bf Q}}
 +\sum\limits_{{\bf Q}_1 \ne {\bf Q}_2}
{\cal U}_{{\bf Q}_1,{\bf Q}_2,{\bf Q}_1,{\bf Q}_2}
-2u\sum\limits_{{\bf Q}}\psi^\dagger_{{\bf Q}}({\bf x})\psi_{{\bf Q}}({\bf x})
\bar n({\bf x}),
\end{eqnarray}
which is that part of the forward scattering term (\ref{H11})---including the 
mean field correction---and affects only the non-condensate band. 

As in QKI we have defined the vector in the space of
$ {\bf M},{\bf N}, $ called    $ {\bf e}_{1234} $  
which can be written
\begin{eqnarray}\Label{eq2.17}
&&{\bf e}_{1234}\Big |_{{\bf Q}_1}={\bf e}_{1234}\Big |_{{\bf Q}_2}
 = -{\bf e}_{1234}\Big |_{{\bf Q}_3}=-{\bf e}_{1234}\Big |_{{\bf Q}_4}=1.
\nonumber \\&&
\end{eqnarray}
with all other components are zero. Thus  $ {\bf e}_{1234} $  represents a 
collision, and we have used the notation $ \sum_{\bf e}$ as a shorthand for a 
summation over $ {\bf Q}_1,{\bf Q}_2,{\bf Q}_3,{\bf Q}_4$.
\subsubsection{Mean-field coupling term}
This is the term which gives rise to mean-field effects, and arises from that 
part of the forward scattering term (\ref{H11}) which mixes together the 
condensate-band and non-condensate band operators.  It is
\begin{eqnarray}\Label{MeanField1}
\dot v_{\bf N}(t)|_{\rm mf} &=& -{i\over\hbar}
[ H_{\rm mf}, v_{\bf N}(t)]
\end{eqnarray}
with
\begin{eqnarray}\Label{MeanField2}
 H_{\rm mf}&=& {2u}\int d^3{\bf x}\, \,
\Bigg\{
\bigg(\sum\limits_{{\bf Q}}
\psi^\dagger_{{\bf Q}}({\bf x})\psi_{{\bf Q}}({\bf x)}-\bar n({\bf x})\bigg)
\bigg(\phi^\dagger({\bf x})\phi({\bf x})-\bar\rho({\bf x})\bigg)
\Bigg\}
\end{eqnarray}
This term is a relatively small coupling term, since the major mean field 
effects have been canceled. For ease of writing, a c-number term 
$ 2\int d^3{\bf x}\, \,\bar n({\bf x})\bar\rho({\bf x})$ has been added, 
leading to the factorized form given.  Of course this does not 
affect the equations of motion.
\subsubsection{Coupling terms}
These terms give all transfer of energy and population 
between $ R_{\rm C}$ and $ R_{\rm NC}$, and are:
\Startsmall
\begin{mathletters}
\begin{eqnarray}\Label{bba21a}
&&\dot v_{\bf N}(t)\Big |_{\rm coupling} =
-{\pi \over  \hbar^2}\int d^3{\bf x}\,\int d^3{\bf x}'\,\Bigg(
2\sum_{432}\sum_m\bigg\{
\delta^p(\Delta\omega_{234}-\epsilon_m)
Z_{432}({\bf x})Z^\dagger_{432}({\bf x'})X^\dagger_m({\bf x})X_m({\bf x}')
v_{\bf N}(t)
\nonumber \\ &&
+\delta^p(-\Delta\omega_{234}+\epsilon_m)
v_{\bf N}(t)
Z_{432}({\bf x})Z^\dagger_{432}({\bf x'})X^\dagger_m({\bf x})X_m({\bf x}')
-2\delta(\Delta\omega_{234}-\epsilon_m)
X_m({\bf x}')Z^\dagger_{432}({\bf x'})
v_{{\bf N}-{\bf e}_{432}}(t)
Z_{432}({\bf x})X^\dagger_m({\bf x})\bigg\}
\nonumber \\ \\  \Label{bba21b}&&  
+2\sum_{432}\sum_m\bigg\{
\delta^p(-\Delta\omega_{234}+\epsilon_m)
Z^\dagger_{432}({\bf x})Z_{432}({\bf x'})X_m({\bf x})X^\dagger_m({\bf x}')
v_{\bf N}(t)
\nonumber \\&&
+\delta^p( \Delta\omega_{234}-\epsilon_m)
v_{\bf N}(t)
Z^\dagger_{432}({\bf x})Z_{432}({\bf x'})X_m({\bf x})X^\dagger_m({\bf x}')
-2\delta(\Delta\omega_{234}-\epsilon_m)
X^\dagger_m({\bf x}')Z_{432}({\bf x'})
v_{{\bf N}+{\bf e}_{432}}(t)
Z^\dagger_{432}({\bf x})X_m({\bf x})\bigg\}
\nonumber \\ &&
 \\ \Label{bba21c} 	   	   	&& 	
+4
\sum_{42}\sum_r\bigg\{\delta^p(\Delta\omega_{24}-\beta_r)
Z_{42}({\bf x})Z^\dagger_{42}({\bf x'})U^\dagger_r({\bf x})U_r({\bf x}')
v_{\bf N}(t)
\nonumber \\&&\qquad
+\delta^p(-\Delta\omega_{24}+\beta_r)
v_{\bf N}(t)
Z_{42}({\bf x})Z^\dagger_{42}({\bf x'})U^\dagger_r({\bf x})U_r({\bf x}')
-2\delta(\Delta\omega_{24}-\beta_r)
U_r({\bf x}')Z^\dagger_{42}({\bf x'})
v_{{\bf N}-{\bf e}_{42}}(t)
Z_{42}({\bf x})U^\dagger_r({\bf x})\bigg\}
 \\&&  	   	   	   	   	   	   	   	   	   	   	
\Label{bba21e} +
\sum_{4}\sum_r\bigg\{\delta^p(\omega_{4}-\beta'_r)
Z_{4}({\bf x})Z^\dagger_{4}({\bf x'})V^\dagger_r({\bf x})V_r({\bf x}')
v_{\bf N}(t)
\nonumber \\&&\qquad
+\delta^p(-\omega_{4}+\beta'_r)
v_{\bf N}(t)
Z_{4}({\bf x})Z^\dagger_{4}({\bf x'})V^\dagger_r({\bf x})V_r({\bf x}')
-2\delta(\omega_{4}-\beta'_r)
V_r({\bf x}')Z^\dagger_{4}({\bf x'})
v_{{\bf N}-{\bf e}_{4}}(t)
Z_{4}({\bf x})V^\dagger_r({\bf x})\bigg\}
 \\&&  	   	   	   	   	   	   	   	   	   	   	
 \Label{bba21f}+
\sum_{4}\sum_r\bigg\{\delta^p(-\omega_{4}+\beta'_r)
Z^\dagger_{4}({\bf x})Z_{4}({\bf x'})V_r({\bf x})V^\dagger_r({\bf x}')
v_{\bf N}(t)
\nonumber \\&&\qquad
+\delta^p(\omega_{4}-\beta'_r)
v_{\bf N}(t)
Z^\dagger_{4}({\bf x})Z_{4}({\bf x'})V_r({\bf x})V^\dagger_r({\bf x}')
-2\delta(\omega_{4}-\beta'_r)
V^\dagger_r({\bf x}')Z_{4}({\bf x'})
v_{{\bf N}+{\bf e}_{4}}(t)
Z^\dagger_{4}({\bf x})V_r({\bf x})\bigg\}\Bigg)
\end{eqnarray}
\end{mathletters}
\Endsmall
\narrowtext
These expressions use the function
\begin{eqnarray}\Label{deltap}
\delta^p(\omega) &=& \delta(\omega) -{i\over\pi}{{\rm P}\over \omega}.
\end{eqnarray}
However, for simplicity we shall usually  neglect the dispersive effects which 
arise from the principal valu integral by 
setting $ \delta^p \to \delta$ everywhere.

\subsubsection{Full master equation}
The full density operator, in the approximation being used for these equations 
of motion, can be written as
\begin{eqnarray}\Label{fullme1}
\rho(t) &=& \sum_{\bf N}v_{\bf N}(t),
\end{eqnarray}
where the summation is over all $ {\bf N}$ in the non-condensate band.  
Summing 
over $ {\bf N}$ in (\ref{bba202}), (\ref{bba203a}--\ref{bba203e}),and in 
(\ref{bba21a}--\ref{bba21f}), it is easy to see that one obtains an 
equation for $ \rho(t)$ which is of exactly the same form, but in which
$v_{\bf N}(t), v_{{\bf N}-{\bf e}}(t)\to \rho(t) $ for any 
$ {\bf N},{\bf e}$.

\section{Interpretation of the master equation}\label{IV}
The master equation gives a representation of transitions due to 
collisions between states, defined by wavelets in the 
non-condensate-band, or eigenfunctions in the condensate band 
in their respective effective potentials which 
already include the majority of mean-field effects.  The collisional 
terms include those purely within the non-condensate-band as in
(\ref{bba203e}), and those which give rise to energy and matter 
transfer between the two bands, as in (\ref{bba21b}--\ref{bba21f}).
In addition to the truly irreversible collisional terms, there are also terms 
which arise from principal value integrals, which cause energy level shifts, 
since they take the form of a commutator.  These terms are explicit in 
(\ref{bba203d}),
and are implicit in the $ \delta^P$ terms in (\ref{bba21b}--\ref{bba21f}), and 
are analogous to the Lamb and Stark shift terms which arise similarly in 
quantum optics.   They are equivalent to energy level shifts found by using 
second order perturbation theory.

\subsection{Stationary solutions of the master equation}
The residual mean-field terms and the level shifts are not expected to be 
large, nor do they in any way change the essential picture of transitions.
To the extent that we can neglect these terms,
the quantum kinetic master equation in the form (\ref{bba201}) conserves 
\begin{eqnarray}\Label{tp23}
H_{a}^{\rm eff}+ H_{\rm C}^{\rm eff} &\equiv& H_{\rm tot}^{\rm eff}
\\
\Label{tp24}
N_{\rm NC} + N_{\rm C} &=& N,
\end{eqnarray} 
so that any function of these provides a stationary solution.  The grand 
canonical stationary solution takes the form
\begin{eqnarray}\Label{tp25}
\rho &\propto& \exp\left(-{H_{a}^{\rm eff}-\mu N_{\rm NC}\over kT}\right)
\otimes\exp\left(-{ H_{\rm C}^{\rm eff}-\mu N_{\rm C}\over kT}\right).
\nonumber \\
\end{eqnarray}
but the fact that $ N_{\rm NC} + N_{\rm C}$ is conserved means that the 
restriction of the grand canonical form to any fixed value of $ N_{\rm NC} + 
N_{\rm C}$ must also be a solution.
The correlation functions of the non-condensate operators are quite 
straightforward
\begin{eqnarray}\Label{tp26}
\langle A^\dagger_{{\bf Q},{\bf r}} A_{{\bf Q}',{\bf r}'}\rangle
& = &
\delta_{{\bf r},{\bf r}'}\delta_{{\bf Q},{\bf Q}'}\bar N_T({\bf Q})
\end{eqnarray}
with 
\begin{eqnarray}\Label{tp27}
&&\bar N_T({\bf Q}) =
\left[ \exp\left({E({\bf Q}) -\mu\over k T}\right) -1\right]^{-1}
\\ \Label{tp 27}
&&\qquad\approx
\left[ 
\exp\left({\hbar^2{\bf K}({\bf Q},{\bf r})^2/2m +
 V_{\rm NC}^{\rm eff}({\bf r}) -\mu\over k T}\right) -1\right]^{-1}.
\nonumber \\
\end{eqnarray}
We can also deduce that the stationary field correlation functions can be 
written
\begin{eqnarray}\Label{tp29}
&&\langle\psi_{\bf Q}^\dagger({\bf x})\psi_{{\bf Q}'}({\bf x}')\rangle
\nonumber\\
&&\qquad
= \delta_{{\bf Q},{\bf Q}'}\bar N_T({\bf Q})
\sum_{\bf r}w_{{\bf K}({\bf Q},{\bf r}),{\bf r}}({\bf x})
           w^*_{{\bf K}({\bf Q},{\bf r}),{\bf r}}({\bf x}')
\\
&&\qquad\Label{tp30}
\approx\delta_{{\bf Q},{\bf Q}'}\bar N_T({\bf Q})
e^{i{\bf K}\left({\bf Q},{{\bf x}+{\bf x}'\over 2}\right) 
\cdot({\bf x}-{\bf x}')}g_{\bf Q}({\bf x},{\bf x}'),
\end{eqnarray}
where the last line comes from the asymptotic form (\ref{in.wv10}).  
(These forms will also be valid to a very good 
degree of approximation even in a canonical ensemble formulation in 
which  $ N_{\rm NC}$ is fixed.)

\subsection{Relationship to the Hartree-Fock-Bogoliubov method}
\label{H-F-B}
The inclusion of the mean-field terms defines the stationary solutions of the 
Hamiltonian part in a form which is closely related to the 
Hartree-Fock-Bogoliubov method 
\cite{HFB}.
The task which must be accomplished here is the construction of best estimates 
of the condensate-band density $ \bar\rho({\bf x})$ and of the 
non-condensate-band density $ \bar n({\bf x})$ as defined in Sect.
\ref{MeanFieldCorr}.

\subsubsection{Non-condensate-band density $ \bar n({\bf x})$}
 The non-condensate-band density must be chosen in the form which takes account 
of the projected form of the interaction in (\ref{H11}), i.e.,
\begin{eqnarray}\Label{mf.a2}
\bar n({\bf x}) &=& \sum_{{\bf Q}\in R_{\rm NC}}
\left\langle\psi_{\bf Q}^\dagger({\bf x})\psi_{\bf Q}({\bf x})\right\rangle
 .
 \end{eqnarray}
and when we are dealing with the stationary state, this is given through 
(\ref{tp30}) by as 
\begin{eqnarray}\Label{HF1}
&& \bar n({\bf x}) = \sum_{{\bf Q}\in R_{\rm NC}}
\bar N_T({\bf Q})g_{\bf Q}({\bf x},{\bf x})
\\ \Label{HF2}
&&\quad\approx\int_{E({\bf K},{\bf x})>E_R} d^3{\bf K}
\left[\exp\left(E({\bf K},{\bf x}) -\mu\over kT\right)-1\right]^{-1}.
\nonumber \\
\end{eqnarray}
Here 
\begin{eqnarray}\Label{HF3}
E({\bf K},{\bf x})&=&
 {\hbar^2{\bf K}^2\over 2m} + V^{\rm eff}_{\rm NC}({\bf x})
\\ \Label{HF4}
&=&  {\hbar^2{\bf K}^2\over 2m} + V_{T}({\bf x})
+2u\big(\bar\rho({\bf x}) +\bar n({\bf x})\big).
\end{eqnarray}
This amounts to a {\em local density approximation} for the atom density of the 
non-condensate-band.

\subsubsection{Condensate-band density $ \bar \rho({\bf x})$}
It is now necessary to specify $ \bar\rho({\bf x})$, which means we need to 
know the condensate state. From (\ref{tp25}) we see that this requires the 
diagonalization of $ H_{\rm C}^{\rm eff}-\mu N_{\rm C} $, where $ H_{\rm C}^{
\rm eff} $
is given by (\ref{mf.302}).  This can be done approximately by using the 
Bogoliubov method \cite{BOG}, generalized to its number conserving form as 
given in 
\cite{trueBog}, but using the $ V^{\rm eff}_{\rm C}({\bf x})$ potential, which 
includes the mean field terms arising from the non-condensate-band.  This would 
mean that we write
\begin{eqnarray}\Label{HF5} 
\phi({\bf x}) &\approx& A\left(\xi_{N_{\rm C}}({\bf x}) 
+{1\over\sqrt{N_{\rm C}+1}}\chi({\bf x})\right),
\end{eqnarray}
where the operator $ A$ is a destruction operator for the number of atoms in 
the condensate-band, so $ A^\dagger A $ has the eigenvalue $ N_{\rm C}$, and
$ \xi_{N_{\rm C}}({\bf x})  $ is a solution of the Gross-Pitaevskii 
equation for $ N_{\rm C}$ particles in the effective potential.
Thus 
\widetext
\begin{eqnarray}\Label{HF6}
-{\hbar^2\nabla^2\over 2m}\xi_{N_{\rm C}}({\bf x}) 
+ V_{\rm C}^{\rm eff}({\bf x}) 
\xi_{N_{\rm C}}({\bf x}) 
+uN_{\rm C}\left|\xi_{N_{\rm C}}({\bf x}) \right|^2\xi_{N_{\rm C}}({\bf x}) 
= \mu_{N_{\rm C}} \xi_{N_{\rm C}}({\bf x}) .
\end{eqnarray}
The residual part $ \chi({\bf x})$ is determined in terms of quasiparticles, 
thus
\begin{eqnarray}\Label{HF7}
\chi({\bf x}) &=& 
\sum_m\left[p_m({\bf x})b_m + q_m({\bf x})b^\dagger_m\right]
\end{eqnarray}
where $ p_m({\bf x}) $ and $ q_m({\bf x})$ diagonalize the residual Hamiltonian
\begin{eqnarray}\Label{HF8}
&&{\cal H}_3 =
-{\hbar^2\over 2m}\int d^3{\bf x}\,\chi^\dagger({\bf x})\nabla^2\chi({\bf x})
+\int d^3{\bf x}\,\chi^\dagger({\bf x})
V_{\rm C}^{\rm eff}({\bf x}) 
\chi({\bf x})
\nonumber \\&&\quad
+ \int d^3{\bf x}\bigg\{
{uN_{\rm C}\over 2}\big(\xi_{N_{\rm C}}({\bf x})\chi^\dagger({\bf x})\big)^2 +
{uN_{\rm C} \over 2}\big(\xi_{N_{\rm C}}^*({\bf x})\chi({\bf x})\big)^2
 +
\chi^\dagger({\bf x})\chi({\bf x})\Big(
{2uN_{\rm C}}\big|\xi_{N_{\rm C}}({\bf x})\big|^2-\mu{_{N_{\rm C}}}\Big)\bigg\}
\nonumber \\&&\quad
-{ uN_{\rm C}\over 2}\int d^3{\bf y}\big|\xi_{N_{\rm C}}({\bf y})\big|^4 .
\end{eqnarray}
\narrowtext
The diagonalization of $ {\cal H}_3$ is equivalent to solving the Bogoliubov-de 
Gennes equations
\cite{B De G} with a potential $V_{\rm C}^{\rm eff}({\bf x}) $, and with an 
upper cutoff energy equal to $ E_R -\mu_{N_{\rm C}}$.  The wavefunctions 
$ p_m({\bf x})$and $ q_m({\bf x})$ are the projections of the resulting 
solutions onto the subspace orthogonal to $ \xi_{N_{\rm C}}({\bf x})$.

This procedure gives solutions with any number $ N_{\rm C}$, $ N_{\rm NC}$ of 
condensate- or non-condensate-band particles. The value of the 
$ \bar\rho({\bf x})$ which arises is then
\begin{eqnarray}\Label{HF9}
\bar\rho({\bf x}) &\approx& N_{\rm C}|\xi_{N_C}({\bf x})|^2 
\nonumber \\ &&
+ \sum_m\left\{n_m|p_{m}({\bf x})|^2 + (n_m+1)|q_{m}({\bf x})|^2\right\}
.
\end{eqnarray}

\subsection{Validity  of the method}
\par\noindent
1. The Bogoliubov approximation used here will be valid when the majority of 
the atoms {\em in the condensate-band} are in fact in the condensate level. 
This is not normally the majority of the atoms in the system, since there 
may be a very large number of atoms in the non-condensate-band, which 
are treated essentially by a local density approximation.

\vskip 5pt\noindent 
2. 	The values of $ \bar n({\bf x})$ and $\bar \rho({\bf x})$ are used then to 
determine the mean field potentials, and hence the appropriate energy levels 
which provide a basis within which the master equation now describes time 
evolution, including transfer of energy and atoms from one band to the other, 
as well as the residual reversible processes.

\vskip 5pt\noindent 
3.  The procedure for determining this basis is essentially equivalent to the 
Hartree-Fock-Bogoliubov method in the Popov approximation, but only in its time 
independent form.  

\vskip 5pt\noindent
4. Note that the chemical potentials and temperatures of the stationary 
solutions of the construction so far used can be determined independently
of each other---it is only when we include the master equation coupling terms 
that the condensate-band and non-condensate-band are forced to evolve to an 
equilibrium in which the chemical potentials and temperatures are the same.  
The relevant equations of motion can then be computed in various degrees of 
simplification, and it is these that should be compared with the time-dependent 
Hartree-Fock-Bogoliubov equations of motion. 

\vskip 5pt\noindent
5. The energy levels, wavefunctions and densities given by this process are to 
be seen as a way of choosing a good basis in terms of which the processes 
described by the master equation (\ref{bba201}) can be described.  As well as 
the irreversible processes, there will be corrections to these which also arise 
from the master equation, and which can be in principle computed.

\section{Time-independent local equilibrium approximation}\Label{V}
In this section we will show how the methodology of QKIII arises.  
We assume that the scattering in $ R_{\rm NC}$ 
is strong, and we are only interested in the behavior of the 
condensate and the modes in $ R_{\rm C}$.  We then factorize the total 
density operator into a quasi-thermalized non-condensate part $ 
\rho_B$, and a density operator $ \rho_{\rm C}(t) $, which describes $ 
R_{\rm C}$.  Thus
\begin{eqnarray}\Label{II4.10}
\rho ( t) =\sum\limits_{{\bf N}}v_{{\bf N}}(t) \to  
 \rho _B \otimes \rho_{\rm C}(t),
\end{eqnarray}
and $ \rho_B$ is of course time-independent.
We  use a {\em local equilibrium} approximation, in which
 $ \rho_B$ is such 
that the averages of the products of field operators in $ R_{\rm NC}$ are 
quasi-thermal, in much the same way as was used in the derivation of the 
Uehling-Uhlenbeck equation in QKI.
Thus we  write, as in (\ref{tp30})
\widetext
\begin{eqnarray}\Label{sim1}
\langle\psi_{\bf Q}^\dagger({\bf x})\psi_{{\bf Q}'}({\bf x}')\rangle
= \delta_{{\bf Q},{\bf Q}'}
e^{i{\bf K}\left({\bf Q},{{\bf x}+{\bf x}'\over 2}\right) 
\cdot({\bf x}-{\bf x}')}g_{\bf Q}({\bf x}-{\bf x}') 
F_{\bf Q}\left({{\bf x}+{\bf x}'\over 2}\right).
\end{eqnarray}
This allows for thermalization locally, in contrast to (\ref{tp30}), for which
$ F_{\bf Q}\left({{\bf x}+{\bf x}'\over 2}\right)
\to \bar N_T({\bf Q}) $
 which give a global thermalization.   We use the Gaussian factorization 
properties of the the averages of the products of more than two field 
operators, so that for example, we can write

\Startsmall
\begin{eqnarray}\Label{sim1a}
\langle\psi_{{\bf Q}_1}^\dagger({\bf x})\psi_{{\bf Q}_2}^\dagger({\bf x})
\psi_{{\bf Q}'_1}({\bf x}')\psi_{{\bf Q}'_2}({\bf x}')\rangle
&=& \left( \delta_{{\bf Q}_1,{\bf Q}_1'} \delta_{{\bf Q}_2,{\bf Q}'_2}
 + \delta_{{\bf Q}_1,{\bf Q}_2'} \delta_{{\bf Q}_2,{\bf Q}'_1}\right)
g_{\bf Q}({\bf x}-{\bf x}') ^2 F_{{\bf Q}_1}\left({{\bf x}+{\bf x}'\over 2}
\right)
F_{{\bf Q}_2}\left({{\bf x}+{\bf x}'\over 2}\right)
\nonumber \\
&&\times
e^{\left\{i{\bf K}\left({\bf Q}_1,{{\bf x}+{\bf x}'\over 2}\right) 
+ i{\bf K}\left({\bf Q}_2,{{\bf x}+{\bf x}'\over 2}\right) \right\}
\cdot({\bf x}-{\bf x}')}.
\end{eqnarray}\Endsmall
\subsection{Master equation}
Tracing over the non-condensate modes, we get
\Startsmall
\begin{mathletters}
\begin{eqnarray}\Label{II4.11a}
\dot \rho _{\rm C}( t)  &=&-{i\over\hbar}   \int d^3{\bf x}\,\Bigg[
\phi^{\dagger}( {\bf x})\left(-{\hbar^2\nabla^2\over2m}
+ V_{\rm C}^{\rm eff}({\bf x}) 
+{1\over 2}u\phi^\dagger({\bf x})\phi({\bf x})\right)
\phi( {\bf x})\,\, , \,\,\rho_{\rm C} \Bigg] 
\\ 
 \Label{II4.11b}+\int d^3{\bf x}\int d^3{\bf x}'\Bigg(
&&
\sum_mG^{(+)}({\bf x}-{\bf x}',\epsilon_m)
\left(2X_m({\bf x})\rho_{\rm C}X^\dagger_m({\bf x'})
-\rho_{\rm C}X^\dagger_m({\bf x'})X_m({\bf x})
-X^\dagger_m({\bf x'})X_m({\bf x})\rho_{\rm C}\right)
\\ \Label{II4.11c}
&+&
\sum_mG^{(-)}({\bf x}-{\bf x}',\epsilon_m)
\left(2X^\dagger_m({\bf x})\rho_{\rm C}X_m({\bf x'})
-\rho_{\rm C}X_m({\bf x'})X^\dagger_m({\bf x})
-X_m({\bf x'})X^\dagger_m({\bf x})\rho_{\rm C}\right)
\\ \Label{II4.11d} 
&+&\sum_rM({\bf x}-{\bf x}',\beta_r)
\left(2U_r({\bf x})\rho_{\rm C}U^\dagger_r({\bf x'})
-\rho_{\rm C}U^\dagger_r({\bf x'})U_r({\bf x})
-U^\dagger_r({\bf x'})U_r({\bf x})\rho_{\rm C}\right)
\\ 	\Label{II4.11f}
&+&\sum_rE^{(+)}({\bf x}-{\bf x}',\beta'_r)
\left(2V_r({\bf x})\rho_{\rm C}V^\dagger_r({\bf x'})
-\rho_{\rm C}V^\dagger_r({\bf x'})V_r({\bf x})
-V^\dagger_r({\bf x'})V_r({\bf x})\rho_{\rm C}\right)
\\ \Label{II4.11g}
&+&
\sum_rE^{(-)}({\bf x}-{\bf x}',\beta'_r)
\left(2V^\dagger_r({\bf x})\rho_{\rm C}V_r({\bf x'})
-\rho_{\rm C}V_r({\bf x'})V^\dagger_r({\bf x})
-V_r({\bf x'})V^\dagger_r({\bf x})\rho_{\rm C}\right)\Bigg)
\end{eqnarray}
\end{mathletters}
in which the quantities $ E^{(\pm ) }$, $ M $, $ G^{(\pm ) } $ are 
given by
\begin{mathletters}
\begin{eqnarray}\Label{II4.12a}
G^{( +) }( {\bf x-x}',\omega) &=&{2\pi\over\hbar^2}{\rm Tr}_B\left\{ \sum
\limits_{{123}}
\delta(\omega_1+\omega_2-\omega_3-\omega) 
Z_{123}^{\dagger }({\bf x}') Z_{123}({\bf x}) \rho _B \right\}   
 \\ \Label{II4.12b}
G^{( -) }( {\bf x-x}',\omega) &=&{2\pi\over\hbar^2}
{\rm Tr}_B\left\{ \sum\limits_{{123}}
\delta(\omega_1+\omega_2-\omega_3-\omega) 
Z_{123}({\bf x}) Z_{123}^{\dagger }({\bf x}') \rho _B \right\} 
 \\ \Label{II4.12c}	
M( {\bf x-x}',\omega) &=&{4\pi\over\hbar^2}
{\rm Tr}_B\left\{ \sum\limits_{{12}}
\delta(\omega_1-\omega_2-\omega) 
Z_{12}^{\dagger }({\bf x}') Z_{12}({\bf x}) \rho _B \right\}   
 \\ \Label{II4.12e} 
E^{( +) }( {\bf x-x}',\omega) &=&{\pi\over\hbar^2}
{\rm Tr}_B\left\{ \sum\limits_{{1}}
\delta(\omega_1-\omega) 
Z_{1}^{\dagger }({\bf x}') Z_{1}({\bf x}) \rho _B \right\}   
 \\ \Label{II4.12f}
E^{( -) }( {\bf x-x}',\omega) &=&{\pi\over\hbar ^2}
{\rm Tr}_B\left\{ \sum\limits_{{1}}
\delta(\omega_1-\omega) 
Z_{1}({\bf x}) Z_{1}^{\dagger }({\bf x}') \rho _B \right\} . 
\end{eqnarray}
\end{mathletters}
\Endsmall\narrowtext
The master equation (\ref{II4.11a}--\ref{II4.11g}) is of the same form as 
given in QKIII, apart from 

\vskip 5pt\noindent
1. The expression of the non-condensate band summations 
in terms of $ {\bf Q}$-bands, rather than the continuum approximation chosen in 
QKIII as integrals over $ {\bf K}$ variables. In this degree of approximation, 
this leads only to a change in the range of $ {\bf K}$ 
as a function of $ {\bf x}$.

\vskip 5pt\noindent
2. The potential defining the $ {\bf Q}$-bands is the  effective potential
$ V_{\rm NC}^{\rm eff}({\bf x})$, which depends on the occupation of the 
condensate-band, rather than simply the trapping potential as in QKIII.

\subsection{Interactions between quasiparticles}
As written in (\ref{bba20201}) and (\ref{II4.11a}) the condensate-band 
Hamiltonian is treated exactly.  Using the Bogoliubov approximation, we can
approximate the condensate-band Hamiltonian term
(\ref{II4.11a}) in the form
\begin{eqnarray}\Label{Q1}
H &=& \sum_{m}\epsilon_m b^\dagger b_m +E_0(N).
\end{eqnarray}
 Although the irreversible terms in (\ref{II4.11a}--\ref{II4.11g}) give  
damping and thermalization, this approximation does omit the effect of the 
higher order terms, which can generate  scattering of the quasiparticles by 
quasiparticles. 
Graham \cite{GrahamFluctuations}
has pointed out that these can be significant, even though the quasiparticle 
density of states is rather low, essentially because the very low energy 
quasiparticles involve very large numbers of atoms.  If this pertains, we have 
three possibilities.
\vskip 5pt\noindent
1)
We can treat the quasiparticles in essentially the same manner as the 
particles in the non-condensate band; that is, we can assume they are 
thermalized, and compute scattering rates.  There is no need, however, to use 
wavelets, since we can simply use the discrete quasiparticle states, of which 
there should be relatively few compared to the non-condensate band states.
This would correspond to Graham's methodology.
\vskip 5pt\noindent
2) It is possible that an exact treatment of the condensate band 
Hamiltonian could be feasible, provided it is realistic to take this as 
consisting of only a relatively few quasiparticle levels. 
Stringari \cite{Stringari excitations} has pointed out that in fact in the case 
of a trap, very few of the levels have any genuine phonon-like character.
 This could be done in combination with 
the use of a stochastic wavefunction method for the interactions with the 
non-condensate band.
\vskip 5pt\noindent
3) As an intermediate form of approximation, it could well be feasible to 
use a positive-P-representation methodology, rather like that used by Drummond 
and Corney \cite{Drummond Corney} for the condensate-band Hamiltonian in a 
quasiparticle basis.

We shall leave further developments of this aspect to another publication

\section{Time dependent local equilibrium approximation}\label{VI}
The local equilibrium approximation, when applied to the quantum-kinetic 
master equation of QKI was shown to give rise to the well known 
Uehling-Uhlenbeck equation \cite{UU}.  A similar method applied to this case 
leads to a description of the coupling of thermalized vapor to the condensate.
Together with the equations of motion for the condensate density operator
as in Sec. III.D.3 of QKIII, or the rate equation approximations to these,
this gives a description of the coupled motion of the 
condensate and non-condensate.  From this description one can take further 
limits to get coupled equations of motion in the hydrodynamic limit, or in the 
weak-collision limit.

To be precise, we define 
\begin{eqnarray}\Label{le1}
f_{\bf K}({\bf x}) &\equiv &F_{{\bf Q}({\bf K},{\bf x})}({\bf x})
\end{eqnarray}
where $ F$ is as defined in (\ref{sim1}).  (This is the same as in QKI).  Using 
the same procedures as in QKI, we can eventually arrive at the equation of 
motion:
\begin{eqnarray}\Label{le2}
{\partial f_{\bf K}({\bf x})  \over \partial t}&=&
\left.{\partial f_{\bf K}({\bf x})  \over \partial t}\right |_{\rm UU}
+\left.{\partial f_{\bf K}({\bf x})  \over \partial t}\right |_{\rm mf}
\nonumber \\ &&
+\left.{\partial f_{\bf K}({\bf x})  \over \partial t}\right |_{1}
+\left.{\partial f_{\bf K}({\bf x})  \over \partial t}\right |_{2}
+\left.{\partial f_{\bf K}({\bf x})  \over \partial t}\right |_{3 }
\end{eqnarray}
Of these, the first two terms represent flow and collisions of atoms
in the non-condensate-band, and the last 
three are irreversible master equation terms, which represent the transfer of 
energy and atoms between the condensate-band and the non-condensate-band.

The equation of motion (\ref{le2})  depends on the condensate-band state 
through the second term, which depends on the mean field determined by the mean 
condensate density $ \rho({\bf x},t)$, as defined by (\ref{le304}).  Thus we 
must complement (\ref{le2}) with an appropriate equation of motion for the 
condensate.  At the simplest, we can simply use rate equations for the 
condensate level occupations
and compute the condensate density from (\ref{HF9}).  The condensate equations 
of motion should 
be written in terms of the {\em time-independent\/} basis set as in 
(\ref{HF7}), which uses the time-independent mean-field potential
$ V_{\rm C}^{\rm eff}({\bf x})$.
The deviation from the average density $ n({\bf x},t)$ induces the additional 
Hamiltonian term (\ref{MeanField1},\ref{MeanField2}), which can also be 
expressed in the Bogoliubov basis.

The process of condensate growth gives rise to an {\em apparent\/} 
time-dependence as follows.  The basis set of states is different for every 
pair of values $ N_{\rm NC}, N_{\rm C}$---thus the set  
$ N_{\rm NC}, N_{\rm C}$
can be regarded as defining a set of independent sectors, each with its own 
basis states.  When the coupling terms between the condensate-band and 
the non-condensate band are taken into account, the effect is to transfer the 
system between sectors by the transitions
$ N_{\rm NC}, N_{\rm C}\to   N_{\rm NC}\pm 1, N_{\rm C}\mp 1$, and to provide a 
new basis for the density operator.  However this can be accounted by a slight 
modification to the Bogoliubov expansion---thus this apparent 
time dependence can in fact be fully accounted for by the existing 
procedures.

\subsection{Collisions and flow in the non-condensate}
\subsubsection{Uehling-Uhlenbeck term}
This term is the part arising from streaming terms and  collisions between 
atoms in the non-condensate-band with each other, 
and takes the form
\widetext
\begin{eqnarray}\Label{le3}
\left.{\partial f_{\bf K}({\bf x})  \over \partial t}\right |_{\rm UU} &=&
 \left({\hbar {\bf K}\cdot \nabla_{{\bf x}}\over m}
-{\nabla_{\bf x}V_{\rm NC}^{\rm eff}({\bf x})\cdot\nabla_{\bf K}\over\hbar}
\right)f_{\bf K}({\bf x})
\nonumber\\
&& +\frac{2| u|^2}{h^2}
\int\limits_{R_{\rm NC}} d^3{\bf K}_2
\int\limits_{R_{\rm NC}} d^3{\bf K}_3
\int\limits_{R_{\rm NC}}d^3{\bf K}_4 
\delta ( {\bf K}+{\bf K}_2-{\bf K}_3-{\bf K}_4)
\delta (\omega +\omega_2-\omega_3-\omega_4)
\nonumber \\
&&\times \Big\{
f_{{\bf K}}( {\bf x}) f_{{\bf K}_2}( {\bf x} )
[ f_{{\bf K}_3}( {\bf x}) +1] [ f_{ {\bf K}_4}( {\bf x}) +1]
 - [f_{{\bf K}}( {\bf x}) +1] [ f_{{\bf K}_2}( {\bf x}) +1]
 f_{{\bf K}_3}({\bf x}) f_{{\bf K}_4}( {\bf x}) \Big\}
\end{eqnarray}
Notice that the range of energies available is restricted to the 
condensate-band, which is consistent with the streaming term---thus there are 
no terms 
here which can transfer particles out of $ R_{\rm NC}$.
\narrowtext
\subsubsection{Mean-field corrections to the Uehling-Uhlenbeck term}
The \UU term includes only the flow of the non-condensate which arises as a 
result of the (time-independent) effective potential 
$ V^{\rm eff}_{\rm NC}=2u\bar\rho({\bf x}) + 2u\bar n({\bf x}) 
+V_T({\bf x})$.
However, when the condensate is not time-independent, the cancellation of the 
average values of $ H_{\rm mf}$ as in (\ref{bba230f},\ref{MeanField2})
does not occur, since 
$ \bar n({\bf x})$ and $ \bar\rho({\bf x})$ are taken as time independent 
estimates
of the mean values.  Including these in the equation of motion for
$ f_{\bf K}({\bf x})$ gives rise to extra terms proportional to 
\begin{eqnarray}\Label{le301}
&&
\left\langle\left[\int d^3{\bf y}
\sum_{\bf Q'}\psi^\dagger_{\bf Q'}({\bf y})\psi_{\bf Q'}({\bf y})\, , \,
\psi^\dagger_{\bf Q}({\bf x})\psi_{\bf Q}({\bf x}) \right ]\right\rangle
\nonumber \\
&&
\times \left\langle
\sum_{\bf Q''}\psi^\dagger_{\bf Q''}({\bf y})\psi_{\bf Q'' }({\bf y})
+\phi^\dagger({\bf y})\phi({\bf y}) -\bar n({\bf y}) -\bar\rho({\bf y})\right
\rangle
.
\nonumber \\
\end{eqnarray}
The commutator is non-zero only for $ {\bf Q}' = {\bf Q}$ since the commutators
$ \psi_{\bf Q}({\bf x})$ etc.\ are non-local.  If we use the form (\ref{sim1})
to evaluate this average this commutator does vanish---a non-vanishing  result 
only appears when we allow corrections to this as a result of the 
non-equilibrium state implicit in the situation.  The detailed working is given 
in Appendix~\ref{AppB}; here we shall merely state the result, which is 
intuitively obvious, that for sufficiently smooth distributions in $ {\bf Q}$
and $ {\bf x}$, we find
\begin{eqnarray}\Label{le302}
&&\left.  {\partial f_{\bf K}({\bf x}) \over\partial t}\right |_{\rm mf}
\nonumber \\
&&\,= 2u\nabla_{\bf x}\left\{
\bar\rho({\bf x},t)-\bar\rho({\bf x})+\bar n({\bf x},t)-\bar n({\bf x})\right\}
\cdot \nabla_{\bf K}f_{\bf K}({\bf x})
\nonumber \\
\end{eqnarray}where $\bar n({\bf x},t)$ and $\bar \rho({\bf x},t)$ are the time dependent 
versions of 
the condensate-band and non-condensate-band densities, i.e.,
\begin{eqnarray}\Label{le303}
\bar n({\bf x},t) &=& 
\sum_{\bf Q'}{\rm Tr}\left(
\psi^\dagger_{\bf Q'}({\bf x})\psi_{\bf Q'}({\bf x})
\bar \rho_B(t)\right)
\\   \Label{le30301}
&=&
{1\over(2\pi)^3}\int_{R_{\rm NC}}d^3{\bf K}\,f_{\bf K}({\bf x}).
\\ \Label{le304}
\bar\rho({\bf x},t) &=& 
{\rm Tr}\left(\phi^\dagger({\bf x})\phi({\bf x})\rho_{\rm C}(t)\right)
.
\end{eqnarray}
Thus the net effect on the \UU term is simply to make the replacement to the 
effective potential
\begin{eqnarray}\Label{le305}
V^{\rm eff}_{\rm NC}({\bf x}) &\to  & V^{\rm eff}_{\rm NC}({\bf x},t)
\nonumber \\ &=&V_T({\bf x}) + 2u\bar\rho({\bf x},t) + 2u\bar n({\bf x},t),
\end{eqnarray}
that is, a time dependent version of the effective potential.


\subsubsection{Condensate growth term}
This is the principal term which gives rise to growth of the condensate. In the 
case that we use the Bogoliubov approximation to the condensate, and the 
notation of Sec.IId of QKIII, to write 
this in terms of the two rate functions
\widetext
\begin{eqnarray}\Label{le4}
{\rm Rate}^{(+)}_{I}({\bf K},{\bf x},\omega) &=& {2u^2\over(2\pi)^2\hbar^2}
\int\limits_{R_{\rm NC}} d^3{\bf K}_2
\int\limits_{R_{\rm NC}} d^3{\bf K}_4
\int d^3{\bf k}\,\delta(\Delta\omega_{2{\bf K}4}({\bf x})-\omega)
\delta({\bf K}+{\bf K}_4 -{\bf K}_2 - {\bf k})
\nonumber \\ && \qquad\qquad\phantom{\int\limits_{R_{\rm NC}}}\times
[1+f_{\bf K}({\bf x})][1+f_4({\bf x})]f_2({\bf x})
{\cal W}_I({\bf x},{\bf k})
\nonumber \\ &-&
{u^2\over(2\pi)^2\hbar^2}
\int\limits_{R_{\rm NC}} d^3{\bf K}_3
\int\limits_{R_{\rm NC}} d^3{\bf K}_4
\int d^3{\bf k}\,\delta(\Delta\omega_{{\bf K}34}({\bf x})-\omega)
\delta({\bf K}_3+{\bf K}_4 -{\bf K} - {\bf k})
\nonumber \\ &&\qquad\qquad\phantom{\int\limits_{R_{\rm NC}}}\times
f_{\bf K}({\bf x})f_2({\bf x})[1+f_3({\bf x})]
{\cal W}_I({\bf x},{\bf k})
\\ \Label{le5}
{\rm Rate}^{(-)}_{I}({\bf K},{\bf x},\omega) &=&{2u^2\over(2\pi)^2\hbar^2}
\int\limits_{R_{\rm NC}} d^3{\bf K}_2
\int\limits_{R_{\rm NC}} d^3{\bf K}_4
\int d^3{\bf k}\,\delta(\Delta\omega_{2{\bf K}4}({\bf x})-\omega)
\delta({\bf K}+{\bf K}_4 -{\bf K}_2 - {\bf k})
\nonumber \\ && \qquad\qquad\phantom{\int\limits_{R_{\rm NC}}}\times
f_{\bf K}({\bf x})f_4({\bf x})[1+f_2({\bf x})]
{\cal W}_I({\bf x},{\bf k})
\nonumber \\ &-&
{u^2\over(2\pi)^2\hbar^2}
\int\limits_{R_{\rm NC}} d^3{\bf K}_3
\int\limits_{R_{\rm NC}} d^3{\bf K}_4
\int d^3{\bf k}\,\delta(\Delta\omega_{{\bf K}34}({\bf x})-\omega)
\delta({\bf K}_3+{\bf K}_4 -{\bf K} - {\bf k})
\nonumber \\ &&\qquad\qquad\phantom{\int\limits_{R_{\rm NC}}}\times
[1+f_{\bf K}({\bf x})][1+f_2({\bf x})]f_3({\bf x})
{\cal W}_I({\bf x},{\bf k})
\end{eqnarray}
The term is then defined as 
\begin{eqnarray}\Label{le6}
\left.{\partial f( {\bf K,x})  \over \partial t}\right |_{1}
&=& {N_{\rm C}}\,{\rm Rate}^{(-)}_{0}({\bf K},{\bf x},\mu_{N_{\rm C}}/\hbar)
-   ({N_{\rm C}}+1){\rm Rate}^{(+)}_{0}({\bf K},{\bf x},\mu_{N_{\rm C}}/\hbar)
\nonumber \\ &&+
\sum_m\Bigg\{
n_m\,{\rm Rate}^{(-)}_{m}({\bf K},{\bf x},(\mu_{N_{\rm C}}+\epsilon_{N_{\rm 
C}}^m)/\hbar)
- (n_m+1){\rm Rate}^{(+)}_{m}({\bf K},{\bf x},(\mu_{N_{\rm C}}+\epsilon_{N_{
\rm 
C}}^m)/\hbar)
\nonumber \\ &&+
(n_m+1){\rm Rate}^{(-)}_{m}({\bf K},{\bf x},(\mu_{N_{\rm C}}-\epsilon_{N_{\rm 
C}}^m)/\hbar)
- n_m\,{\rm Rate}^{(+)}_{m}({\bf K},{\bf x},(\mu_{N_{\rm C}}-\epsilon_{N_{\rm 
C}}^m)/\hbar)
\Bigg\}
\end{eqnarray}
\narrowtext
\subsubsection{Scattering terms}

We also need an abbreviated notation for the number operators of the 
condensate-band
\begin{eqnarray}\Label{comp9}
\tilde n_I &=& \langle S^\dagger_I S_I\rangle
\\  \Label{comp10}
\tilde m_I &=& \langle S_I S_I^\dagger\rangle ,
\end{eqnarray}
and in particular this means
\begin{eqnarray}\Label{comp11}
\tilde n_0 &=& {N_{\rm C}}
\\ \Label{comp12}
\tilde m_0 &=& {N_{\rm C}}+1
\\ \Label{comp13}
\tilde n_m &=& {N_{\rm C}} n_m
\\ \Label{comp14}
\tilde m_m &=& ({N_{\rm C}}+1)(n_m+1)
\\ \Label{comp15}
\tilde n_{-m}&=& {N_{\rm C}} (n_{m}+1)
\\ \Label{comp16}
\tilde m_{-m} &=& ({N_{\rm C}}+1)n_{m}
\end{eqnarray}

Here we can again define some rate functions in the form
\widetext
\begin{eqnarray}\Label{le7}
{\rm Rate}_{IJ}({\bf K},{\bf x},\omega) &=& {8\pi u^2\over\hbar^2}
\int\limits_{R_{\rm NC}} d^3{\bf K}_2
\int d^3{\bf k}\int d^3{\bf k}'
\,\delta(\Delta\omega_{2{\bf K}}({\bf x})-\omega)
\delta({\bf K} -{\bf K}_2 - {\bf k}+{\bf k}')
\nonumber \\ && \qquad\qquad\phantom{\int\limits_{R_{\rm NC}}}\times
[1+f_{\bf K}({\bf x})]f_2({\bf x})
{\cal W}_I({\bf x},{\bf k}){\cal W}_J({\bf x},{\bf k}')
\nonumber \\ &-&
{8\pi u^2\over\hbar^2}
\int\limits_{R_{\rm NC}} d^3{\bf K}_4
\int d^3{\bf k}\int d^3{\bf k}'
\,\delta(\Delta\omega_{{\bf K}4}({\bf x})-\omega)
\delta({\bf K}_4-{\bf K} - {\bf k}+{\bf k}')
\nonumber \\ &&\qquad\qquad\phantom{\int\limits_{R_{\rm NC}}}\times
f_{\bf K}({\bf x})[1+f_4({\bf x})]
{\cal W}_I({\bf x},{\bf k}){\cal W}_J({\bf x},{\bf k}')
\end{eqnarray}
The term is then defined as
\begin{eqnarray}\Label{701}
\left.{\partial f( {\bf K,x})  \over \partial t}\right |_{2}
&=& \sum_{IJ}{\rm Rate}_{IJ}\big({\bf K},{\bf x},\Omega_{I}({N_{\rm C}})-
\Omega_{J}({N_{\rm C}}) \big)
\tilde n_I\tilde m_J
\end{eqnarray}

\subsubsection{Three-quasiparticle terms}
\begin{eqnarray}\Label{le8}
{\rm Rate}^{(+)}_{IJK}({\bf K},{\bf x},\omega) &=& { 2\pi u^2\over\hbar^2}
\int d^3{\bf k}\int d^3{\bf k}'\int d^3{\bf k}''
\,\delta(\Delta\omega_{{\bf K}}({\bf x})-\omega)
\delta({\bf K} - {\bf k}+{\bf k}'+{\bf k}')
\nonumber \\ && \qquad\qquad\phantom{\int\limits_{R_{\rm NC}}}\times
[1+f_{\bf K}({\bf x})]
{\cal W}_I({\bf x},{\bf k})
{\cal W}_J({\bf x},{\bf k}')
{\cal W}_K({\bf x},{\bf k}')
\\
{\rm Rate}^{(-)}_{IJK}({\bf K},{\bf x},\omega) &=& { 2\pi u^2\over\hbar^2}
\int d^3{\bf k}\int d^3{\bf k}'\int d^3{\bf k}''
\,\delta(\Delta\omega_{{\bf K}}({\bf x})-\omega)
\delta({\bf K} - {\bf k}+{\bf k}'+{\bf k}')
\nonumber \\ && \qquad\qquad\phantom{\int\limits_{R_{\rm NC}}}\times
f_{\bf K}({\bf x})
{\cal W}_I({\bf x},{\bf k})
{\cal W}_J({\bf x},{\bf k}')
{\cal W}_K({\bf x},{\bf k}')
\end{eqnarray}
The term is then defined as
\begin{eqnarray}\Label{le9}
\left.{\partial f( {\bf K,x})  \over \partial t}\right |_{3}
&=& \sum_{IJK}\Big\{{\rm Rate}^{(+)}_{IJ}
\left[{\bf K},{\bf x},\Omega_{K}({N_{\rm C}})+\Omega_{I}({N_{\rm C}})-
\Omega_{J}({N_{\rm C}})\right]
\tilde n_I\tilde m_J\tilde m_K
\nonumber \\ &&
+
{\rm Rate}^{(-)}_{IJK}
\left[{\bf K},{\bf x},\Omega_{K}({N_{\rm C}})+\Omega_{I}({N_{\rm C}})-
\Omega_{J}({N_{\rm C}})\right]
\tilde m_I\tilde n_J\tilde n_K
\Big\} .
\end{eqnarray}
\narrowtext
The various rate in the growth, scattering, and three-$ \phi$ terms are 
compatible with the same rates as seen by the condensate, i.e., 
equations (126,127,141,143,144) of QKIII,
in the sense 
that the rate at which a process removes/adds a particle to the condensate-band
matches exactly with the corresponding rate at which the process adds/removes a 
particle from the non-condensate-band.

\section{Conclusion}\label{VII}
The principal results, in the sense that they can be applied, are the master 
equations of the time-independent local equilibrium approximation, 
Sect.\ref{III}, and of the time-dependent local-equilibrium approximation of 
Sect.\ref{V}.  The first of these was derived in QKIII, but without the 
inclusion of mean-field effects. These are now included by means of the use of 
effective potentials, which account for the majority of such effects, giving a 
description which is quite realistic.  The use of this picture---with some 
technical simplifications---gives the basis for the condensate growth 
methodology of \cite{NewestBosGro}, and this seems to agree quite well with 
experiment.

However, the assumption of a static non-condensate-band is not always 
reasonable, so the methodology of Sect.\ref{V} to get the time dependent local 
equilibrium approximation is necessary.  Fortunately, the basic modification is 
essentially straightforward. The equations are modified so that the 
non-condensate band evolves according to a quantum Boltzmann equation, and the 
content of Sect.\ref{V} is how one includes from the point of view of the 
non-condensate-band equation of motion the correct way of treating the transfer 
of atoms and energy between it and the condensate band.  The necessary 
modifications are given in (\ref{le6},\ref{701},\ref{le9}), are easy to 
understand, but 
have a complexity which arises from the mixture of  {\em particle }creation and 
destruction operators which defines the {\em quasiparticle} creation 
operators---this complexity is intrinsic to the problem and therefore 
unavoidable.

The whole procedure has been developed in terms of the Bogoliubov picture of 
condensate-band atoms moving in the effective potential, which includes 
mean-field effects, and is essentially a kind of Hartree-Fock picture.  
Even though no {\em explicit} depiction of the scattering of quasiparticles by 
each other is given, this can be treated if required by one of the following 
options.

\vskip 5pt\noindent
1. We can ignore these scattering effects entirely, when they are negligible 
compared to the scattering of quasiparticle by non-condensate-band atoms. This 
will certainly be the case if there is a very large proportion of thermal 
vapor.

\vskip 5pt\noindent
2.   Notice however that the condensate band Hamiltonian 
$ H_{\rm C}^{\rm eff}$
as defined in (\ref{bba202})is {\em never} itself explicitly approximated by 
the Bogoliubov form in the master equations.  It is only the transition rates 
induced 
by collisions which have been expressed using the Bogoliubov approximation.  
One could in principle solve the master equation using the full 
$ H_{\rm C}^{\rm eff}$, and the nonlinear mixing induced by the nonlinearity 
must be 
equivalent to the collisions of quasiparticles, because of the presence of the 
randomization induced by the collisions with the vapor.

\vskip 5pt\noindent
3. What seems simplest is a compromise---approximate $ H_{\rm C}^{\rm eff}$ by 
the 
Bogoliubov form taken to the next highest order, which gives rise to Belyayev
terms.  In the $ 1/\sqrt{N}$ expansion in which the Bogoliubov method is valid 
the next higher order terms must be quite small.  This is possible because the 
the condensate band consists only of a small fraction of the levels available, 
and the quartic terms which appear non-negligible in a conventional Bogoliubov
\cite{Girardeau} 
method are taken account of her by the use of the effective potentials.
\vskip 5pt\noindent
Thus the quantum kinetic methodology is complete.  It provides a logical 
combination of kinetics, as given by the quantum Boltzmann equation, with the 
field aspect, as given by the Bogoliubov picture.  It is likely to be as 
accurate as any experiment will require, and has already been used to give a  
quantitative picture of condensate growth \cite{BosGro,NewestBosGro}, 
consistent with experiment.

\appendix
\section{Construction of wavelets}\label{AppA}
\subsection{One dimensional systems}
In this case the wavelets can be 
expressed directly in terms of trap energy eigenfunctions.
Consider the one dimensional trapping Hamiltonian
\begin{eqnarray}\Label{wv1}
H_T &=& {p^2\over 2M} + V(x) 
\\
&=& \sum_n E_n|n\rangle\langle n|
\end{eqnarray}
We relabel these states by dividing them into {\em groups\/} labeled by an 
index 
$ N=0,1,2,\dots ,$ so that there are $ \Delta_N $ states in each group, and we 
can now label the states for a fixed $ N$ by a subsidiary index $ l$.  
Thevalue of the parameter $ \Delta_N$ will be fixed in due course, to give a 
description of maximum convenience.

This means we write the states as $ |N,l\rangle_g$, where
 $ l= 1,2,3,\dots,\Delta_N-1$, and now the 
energy of a given state is 
$ E_{n(N,l)} $, with
\begin{eqnarray}\Label{wv2}
 n(N,l) = l-1+\sum_{N'=0}^{N-1}\Delta_{N'}.
\end{eqnarray}
Using this relabeling, we define the basic wavelet states as
\begin{eqnarray}\Label{wv3}
|N,m\rangle &=& {e^{-\pi i(\Delta_N+1)}\over\sqrt{\Delta_N}}
\sum_{l=1}^{\Delta_N}
e^{2\pi i l m/\Delta_N } |N,l\rangle_g ,
\\
\mbox{where } m &=& 1,2,3,\dots,\Delta_N .
\end{eqnarray}

\Beginfigure
\Label{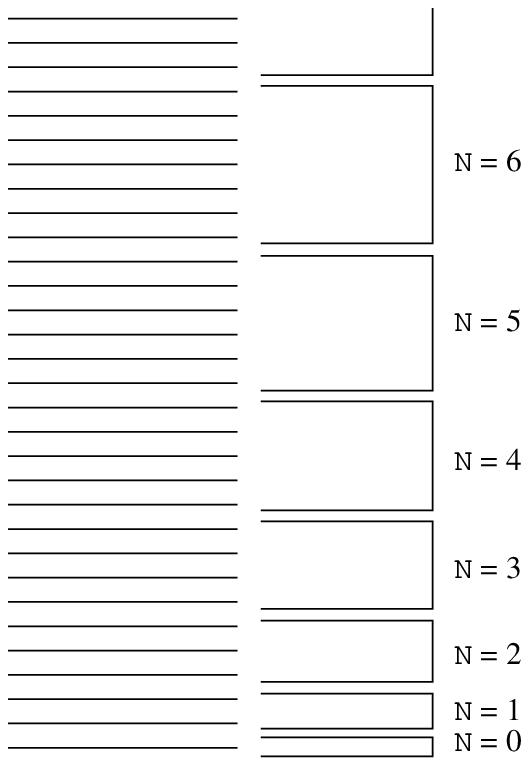}
\epsfig{file=fig2.eps, width=8cm}
\Caption{Fig.2. Grouping of Trap levels.  Wavelets are formed by 
linear combinations of levels in each group.}
\Endfigure

These states form an orthonormal set
\begin{eqnarray}\Label{wv5}
\langle N',m'|N,m\rangle&=& \delta_{NN'}\delta_{mm'}.
\end{eqnarray}
The actual wavelet functions are given by the position space representations of 
these states;
\begin{eqnarray}\Label{wv6}
w_{N,m}(x) &=& \langle x| N,m\rangle
\end{eqnarray}
\subsubsection{Phase space variable $ \theta$}
The best physical variable to use here is
\begin{eqnarray}\Label{wv601}
\theta &\equiv& 2\pi m/\Delta_N.
\end{eqnarray}
This variable is an angle with the range
\begin{eqnarray}\Label{wv602}
\theta &=& 
{2\pi\over\Delta_N}\, , \,
{2\times2\pi\over\Delta_N}\, , \,
{3\times2\pi\over\Delta_N}\, , \,
\dots,
2\pi.
\end{eqnarray}
This range approaches a continuum as $ \Delta_N \to \infty$.

\subsubsection{Expression of the trapping Hamiltonian}
The average energy of the states $ |N,l\rangle$  in the energy band $ N$ is
\begin{eqnarray}\Label{wv7}
\bar E_N &=& {1\over\Delta_N}\sum_{l=1}^{\Delta_N}E(N,l).
\end{eqnarray}
We can now write the trapping
Hamiltonian $ H_T$ as a part diagonal in 
both $ N$ and $ \theta$, and a residue which is diagonal in $ N$ only;
\begin{eqnarray}\Label{wv8}
&&\langle N,\theta |H_T |N',\theta'\rangle 
\nonumber \\  && \qquad
= \delta_{N,N'}\Big\{ \bar E_N\delta_{\theta\theta'}
+\left[
\langle N,\theta |H_T |N',\theta'\rangle -\bar E_N \delta_{\theta\theta'}
\right]
\Big\}\nonumber \\
\end{eqnarray}
It is now necessary to make some approximations in order to get some idea of 
the kinds of functions which will turn up.  The major approximation is to 
assume the energy of each level is approximately a linear function of $ l$ for 
any $ N$.  This is  exact in the case of a harmonic potential.

Thus we write
\begin{eqnarray}\Label{wv9}
E(N,l) &=& \bar E(N) + \epsilon_N(l)
\end{eqnarray}
where now, we approximate
\begin{eqnarray}\Label{wv10}
 \epsilon_N(l) &\approx& \hbar\omega_N\left(l -{\Delta_N +1\over 2}\right), 
\end{eqnarray}
and this equation defines $ \omega_N$.
After some manipulation we find that
\begin{eqnarray}\Label{wv11}
\langle N,\theta |H_T| N',\theta'\rangle &=& 
\delta_{N,N'}\left\{ \bar E_N \delta_{\theta\theta'}
-i{\hbar\omega_N} h_N(\theta'-\theta)\right\}.
\nonumber \\
\end{eqnarray}
Here $ h_N(x)$ is defined by
\begin{eqnarray}\Label{wv12}
h_N(x) &=& 
{\partial \over\partial x }
\left\{\sin\left(\Delta_N x/ 2\right)
\over\Delta_N \sin\left( x/2\right)\right\}.
\end{eqnarray}
\subsubsection{Discrete delta-function approximations}
\label{app-delta}
When $x$ is an integer, the function
\begin{eqnarray}\Label{wv1201}
k_N(x) &=& 
\sin\left(\Delta_N x/ 2\right)\over\Delta_N \sin\left( x/2\right)
\end{eqnarray}
is a Kronecker delta function $ \delta_{x,0}$.
Correspondingly, when $ h_N(x)$ is used in a summation like
\begin{eqnarray}\Label{wv1202}
F_h(x) &=& \sum_{x'}h_N(x-x')F(x'),
\end{eqnarray}
provided $ F(x')$ changes sufficiently smoothly as 
$ x' = \dots,x-2,x-1,x,x+1,x+2,\dots$, then we can write
\begin{eqnarray}\Label{wv1203}
F_h(x) &\approx& {F(x-1)-F(x+1)\over 2},
\end{eqnarray}
and thus $ h_N(x)$ behaves like a ``derivative'' of a Kronecker delta function.
If $ F(x)$ is smooth over a wide range of integer values of $ x$, so that the 
range of $ x$ can be regarded as a continuum, the finite difference 
(\ref{wv1203}) becomes essentially the same as a derivative.  This is the limit 
in which we shall use these functions.

\subsubsection{Phase space behavior of the wavelets}
By using the WKB approximation, we can derive an approximate form for the 
wavelet functions in the form
\begin{eqnarray}\Label{wv13}
w_{\bar E_N,\theta}(x) &\approx&
{1\over\sqrt{p_{\bar E_N}(x)\,\Delta_N}}
\exp\left({i\over\hbar}\int_0^x dx'\,p_{\bar E_N}(x')\right)
\nonumber \\ &&\times
{\sin\{\Delta_N[\theta+\theta(x,\bar E_N)]/2\}\over
\sin\{[\theta+\theta(x,\bar E_N)]/2\}}.
\end{eqnarray}
In this equation, the function
\begin{eqnarray}\Label{wv14}
p_{\bar E_N}(x) &=& \sqrt{2M\big(\bar E_N -V(x)\big)}
\end{eqnarray}
is the classical momentum of the atom at point $ x$, while the function 
\begin{eqnarray}\Label{wv15}
\theta(x,\bar E_N) = M\omega_N\int_0^x{ dx'\over p_{\bar 
E_N}(x')}
\end{eqnarray}
is proportional to the classical time taken for the atom to reach the point 
$ x$; indeed if the trap is exactly harmonic with trap frequency $ \omega$, 
then $ \hbar\omega_N = \hbar\omega$,    and $ \theta(x,\bar E_N)$ is exactly 
the phase angle $ \omega t$ of the oscillator at the time taken to reach the 
point $ x$.

The approximate form of the wavelet function (\ref{wv13}) shows that it is 
sharply peaked when $ \theta = -\theta(x,\bar E_N)$, so that the wavelets 
represent a localization at the point where a classical particle would have 
phase equal to $ \theta(x,\bar E_N)$

\Beginfigure\Label{fig.cells.eps}
\epsfig{file=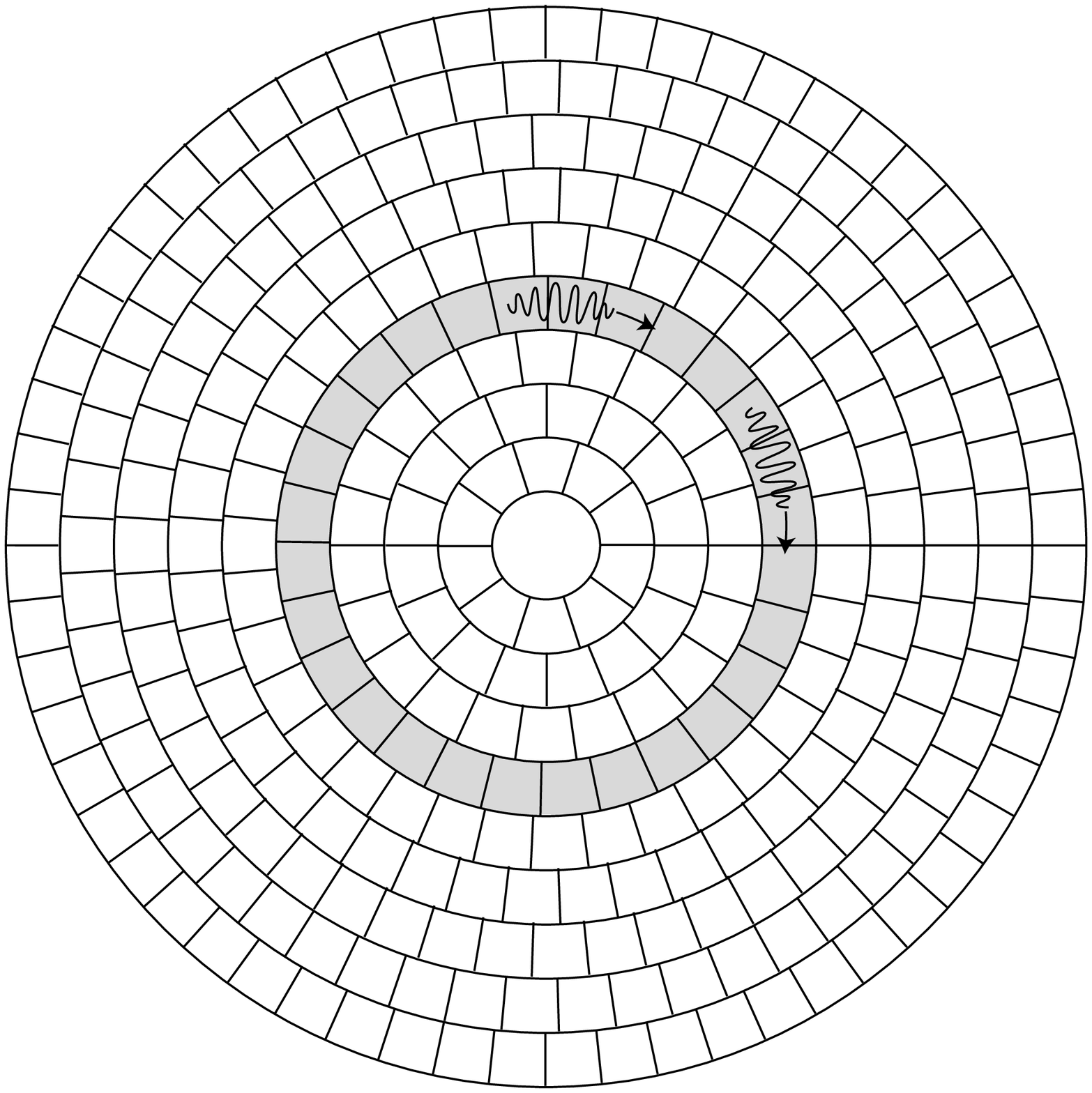, width=8cm}
\Caption{Fig.3. The phase space cells for a harmonic oscillator}
\Endfigure

\subsubsection{Harmonic trapping potential}
In the case of the harmonic oscillator potential $ V(x)= kx^2/2$, we can  
label the wavelets by a 
momentum $ p(N,\theta)$ and position $ x(N,\theta)$ given by
\begin{eqnarray}\Label{wv16}
r(N,\theta) &=& \sqrt{2\bar E_N/ k}\,\cos\theta
\\
\Label{wv17}
p(N,\theta) &=& \sqrt{2M\bar E_N}\,\sin\theta
\end{eqnarray}
and hence get a set of values of $ r$ and $ p $ which can be used to label the 
wavelets.  This set is completely equivalent to the $ N,\theta$ labelling.
\subsubsection{Size of the phase space cells}
The value of $ \Delta_N$ is not yet fixed, and it should be chosen for the 
maximum convenience.  The best choice is one which makes the phase space cells 
all have approximately the same dimensions. For large $ N$, we will show that 
this can be achieved by the choice
\begin{eqnarray}\Label{wv18}
\Delta_N &\approx & \Delta\sqrt{\bar E_N},
\\ \Label{wv19}
\Delta &=& 2\sqrt{2\pi\over\hbar\omega}.
\end{eqnarray}
With this choice, the uncertainties of the phase and energy variables are
\begin{eqnarray}\Label{wv20}
\delta \bar E_N &=& \hbar\omega\Delta\sqrt{\bar E_N}
\\ \Label{wv21}
\delta \theta_N &=& {2\pi\over\Delta\sqrt{\bar E_N}}
\end{eqnarray}
so that, using (\ref{wv16},\ref{wv17}), the uncertainties in $ r$ and $ p$
are
\begin{eqnarray}\Label{wv22}
\delta r &=& 
\left |\sin\theta\right|
{2\pi\sqrt{2}\over \Delta\sqrt{k}}
+\left |\cos\theta\right |
{\hbar\omega\Delta\over 2\sqrt{2k}}
\\ \Label{wv23}
\delta p &=& 
\left|\cos\theta\right |
{2\pi\sqrt{2M}\over \Delta}
+\left|\sin\theta\right |
{\hbar\omega\sqrt{M}\over 2\sqrt{2}}.
\end{eqnarray}
Here $ \delta r$ and $ \delta p$ vary quite strongly with 
the 
phase $ \theta$, unless the coefficients of the $ \sin \theta$ and
 $ \cos\theta$ in each of these equations are the same, and this happens when 
$ \Delta = 2\sqrt{2\pi/\hbar\omega}$, as in (\ref{wv19}).

In this case the uncertainties become
\begin{eqnarray}\Label{wv24}
\delta r &=& {\sqrt{\pi\hbar\over\eta}}\big(|\sin\theta | +|\cos\theta |\big)
\\ \Label{wv25}
\delta p &=& {\sqrt{\pi\hbar\eta}}\big(|\sin\theta | +|\cos\theta |\big).
\end{eqnarray}
with $ \eta =(kM)^{1/4}$.

The coefficients  of $|\sin\theta | +|\cos\theta | $ are essentially the 
variances 
of $ r$ and $ p$ in the ground state of a harmonic oscillator---thus we find
$h/2\le \delta r\,\delta p \le h$, which means that the cells correspond 
basically to minimum uncertainty states.  The ``cells'' are thus square when 
measured in these natural units, and the variation in $ \delta r$ and 
$ \delta p$ arises solely because their orientations depend on $ \theta $.

The choice (\ref{wv18}) of $ \Delta_N$ naturally leads to the the fact that 
for 
large $ N$
\begin{eqnarray}\Label{wv26}
\bar E_N \approx {\Delta^2N^2\over 4}
\end{eqnarray}

\subsection{One dimensional field Hamiltonian in terms of wavelets}
The field Hamiltonian can be written in terms of the non-interacting and the 
collisional parts
\begin{eqnarray}\Label{wv2601}
H &=& H_{\rm kin} + H_{\rm T} + H_{\rm I}.
\end{eqnarray}
We can write the non-interacting part as 
\begin{eqnarray}\Label{wv27}
H_{\rm kin} + H_{\rm T} &=& 
\sum_{N\theta}\bar  E_N a^\dagger_{N\theta} a_{N\theta}
+
\sum_{N\theta\theta'}M_N(\theta,\theta') a^\dagger_{N\theta} a_{N\theta'}.
\nonumber \\
\end{eqnarray}
Here $ M_N(\theta-\theta') $ is defined in terms of the matrix elements
(\ref{wv11},\ref{wv12}) as 
\begin{eqnarray}\Label{wv28}
 M_N(\theta-\theta') =
-i{\hbar\omega_N}h_N(\theta-\theta').
\end{eqnarray}
\subsubsection{Phase space description}
If we use the labelling in terms of $ p$ and $ r$, the structure of the 
description is unaltered, but the appearance of the expression looks 
different.  
Thus we can write
\begin{eqnarray}\Label{wv29}
H_{\rm kin} + H_{\rm T} &=& 
\sum_{rp}\bar E_{rp}a^\dagger_{rp} a_{rp}
+
\sum_{rr'pp'}M(rr'pp') a^\dagger_{rp} a_{r'p'}.
\nonumber \\
\end{eqnarray}
This expression conceals the nature of the wavelet basis, which is implied only 
by the fact that the phase space points $ r,p$ represent wavelets through the 
definition (\ref{wv16},\ref{wv17}), which arranges the points on discrete 
energy shells at positions on these shells determined by the values of 
$ \theta$.  The nature of the function $ M(rr'pp') $ means that it is
nonzero only for points $ r,p$ and $ r',p'$ which are on the same energy shell.
When acting on sufficiently smooth functions we can use the approximate 
relations of sect.\ref{app-delta}, and in particular (\ref{wv1203}), $ F(r,p)$, and  thus approximate
\begin{eqnarray}\Label{wv30}
\sum_{rr'pp'}M(rr'pp')F(r',p') &\approx& 
i\hbar\left({p\over M}{\partial \over\partial r}
-kr{\partial \over\partial p }\right)F(r,p).
\nonumber \\
\end{eqnarray}
Thus the Liouvillian flow arises for sufficiently smooth functions, and we can 
formally write
\begin{eqnarray}\Label{wv31}
[H,a_{rp}] &\approx& {\bar E_{rp}}a_{rp} 
+
i\hbar\left({p\over M}{\partial \over\partial r}
-kr{\partial \over\partial p }\right)a_{rp},
\end{eqnarray}
provided $ a_{rp}$ is interpreted as being evaluated
in an average over a density operator for a state which satisfies the 
smoothness requirement.

\subsection{Three dimensional formulation}
The case of a harmonic potential is particularly simple to treat, since the 
energies in the three different dimensions are separately conserved. 
We will use a notation in terms of these three energy components as follows.
Each ``cell'' is specified by a set of six quantities 
$ (N_1,N_2,N_3, \theta_1, \theta_2, \theta_3)$, from which we can form the two 
vector quantities $ {\bf r},{\bf Q}$, defined by
\begin{eqnarray}\Label{wv32}
r_i &=& \sqrt{2\bar E_{N,i}\over k}\cos\theta_i,
\\
\hbar Q_i &=& \sqrt{2M\bar E_{N,i}}\,{\rm sign}\left(\sin\theta_i\right).
\end{eqnarray}
The definition of $ {\bf Q}$ ensures that the set $ {\bf r},{\bf Q}$ do 
determine $ (N_1,N_2,N_3, \theta_1, \theta_2, \theta_3)$ uniquely; the sign of 
$ Q_i$ determines the sign of the $ i$-component of the momentum, since for a 
given $ E_i$ and $ r_i$ this can have two possible directions.  
In terms of these variables, we now use the notation 
$ w_{\bf Q}({\bf x},{\bf r})$ for the wavelet function, and the notation 
$ A_{\bf Q,r}$ for the corresponding destruction operator, so that the 
field operator can be expressed as 
\begin{eqnarray}\Label{wv33}
\psi({\bf x}) &=& \sum_{\bf Q,r}w_{\bf Q}({\bf x},{\bf r}) A_{\bf Q,r}
\\
&=& \sum_{\bf Q}\psi_{\bf Q}({\bf x})
\end{eqnarray}
The commutation relations are of course
\begin{eqnarray}\Label{wv34}
[ A_{\bf Q,r}, A^\dagger_{\bf Q',r'}]= \delta_{\bf rr'}\delta_{\bf QQ'}
\end{eqnarray}
The commutation relations for the $ \psi_{\bf Q}({\bf x})$ will be needed; they 
are
\begin{eqnarray}\Label{wv3301}
&&
 [\psi_{\bf Q}({\bf x}), \psi^\dagger_{\bf Q'}({\bf x'})] = \delta_{\bf QQ'}
\sum_{\bf r}w^*_{\bf Q}({\bf x},{\bf r})w_{\bf Q}({\bf x'},{\bf r})
\\ 
&&\qquad\approx
\delta_{{\bf Q},{\bf Q}'}
e^{i{\bf K}\left({\bf Q},{{\bf x}+{\bf x}'\over 2}\right) 
\cdot({\bf x}-{\bf x}')}g_{\bf Q}({\bf x},{\bf x}').
\end{eqnarray}
The approximation to get the second line comes from the WKB form, 
(\ref{wv13}).
\subsubsection{General potentials}
The construction here is specific to a potentials which can be written as the 
sum of $ x$, $ y$ and $ z$ dependent parts, of which the three-dimensional 
harmonic potential is the most useful example.  Even though harmonic traps are 
most 
commonly used, the effect of the condensate on the other particles acts like 
an 
additional potential, and it will therefore be necessary for us to consider 
more general 
potentials.  We will, without proof, assume that it is always possible to find 
an appropriate wavelet basis of the kind set up here.  Although this may not 
be 
so, particularly in the case of non-integrable systems, it seems very likely 
that the results of our analysis will continue to be true.  we say this 
because  
our methodology really depends only on a local properties, and almost every 
potential we are likely to use is locally harmonic.

\section{Time dependent mean-field terms}\label{AppB}
We will show that, under the requirement that that the momentum disrtibution 
function is smooth, we may use the approximate expression
\begin{eqnarray}\Label{AppB.1}
 \langle\psi^\dagger_{\bf Q}({\bf x})
\psi^\dagger_{\bf Q}({\bf x}') \rangle
&=&e^{i{\bf K}({\bf Q},{\bf u})\cdot{\bf v}}\bigg\{
g({\bf x},{\bf x}') F\big({\bf K}({\bf Q},{\bf u}),{\bf u}\big)
\nonumber \\
&&-i\big(\nabla_{\bf v} g({\bf x},{\bf x}') \big)\cdot
\nabla_{\bf K}F\big({\bf K}({\bf Q},{\bf u}),{\bf u}\big)\bigg\}
\nonumber \\
\end{eqnarray}
in which $ {\bf x}={\bf u}+{\bf v}$ and $ {\bf x}'={\bf u}-{\bf v}$.
We will do this explicitly for the case of no trapping potential as follows.  
In this case we consider 
\begin{eqnarray}\Label{AppB.2}
&& \langle\psi^\dagger_{\bf K}({\bf x})
\psi^\dagger_{\bf K}({\bf x}') \rangle
\nonumber \\
&&\qquad ={1\over(2\pi)^3}\int\!\!\!\int^{\bf K +\Delta}_{\bf K-\Delta}d^3{\bf 
k}\,d^3{
\bf k}'\,
\langle a_{\bf k}^\dagger a_{\bf k'}\rangle 
e^{i({\bf k}\cdot{\bf x}-{\bf k}'\cdot{\bf x}')}
\end{eqnarray}
We can now express the smoothness requirement as the condition that the mean
$ \langle a_{\bf k}^\dagger a_{\bf k'}\rangle$ can be expressed as a function 
$ R({\bf p},{\bf q})$
of $ {\bf p}= ({\bf k}+{\bf k}')/2$ and $  {\bf q}={\bf k}-{\bf k}'$. 
The requirement that the different bands in $ {\bf K}$ be uncorrelated, which 
is fundamental to the method,  only makes physical sense if any correlation 
that does exist between $ {\bf k}$ and $ {\bf k}'$ is over a range in 
$ {\bf k}$ that is very much less than $ \Delta$, the spacing between the 
bands. Thus $ R({\bf p},{\bf q})$ must be sharply peaked in $ {\bf q}$. On the 
other hand, the dependence of $ R({\bf p},{\bf q})$ on $ {\bf p}$ represents 
the overall momentum distribution, which should be smooth over the range 
$ \Delta$ between the bands.

This means 
that we can change to the variables $ {\bf p},{\bf q}$ in the integral, and to 
a good approximation let the limits of the $ {\bf q}$ integration go to 
infinity, while those of the $ {\bf p}$ are 
$ {\bf K}-{\bf \Delta},{\bf K}+{\bf \Delta}$.  We implement the smoothness in $ 
{\bf p}$ requirement by using a Taylor expansion for $ R({\bf p},{\bf p})$, 
getting first order approximation
\begin{eqnarray}\Label{AppB.3}
&& \langle\psi^\dagger_{\bf K}({\bf x})\psi^\dagger_{\bf K}({\bf x}') \rangle
\nonumber \\
&&\approx {1\over(2\pi)^3}
\int_{-\infty}^\infty\!\!\!\!\!\! d^3{\bf q}
\int^{\bf K +\Delta}_{\bf K-\Delta}\!\!\!\!\!\!\!\!d^3{\bf p}\,
e^{-i({\bf p}\cdot{\bf u}+{\bf q}\cdot{\bf v})}
\nonumber \\ &&\qquad\qquad\qquad\times
\bigg\{  R({\bf K},{\bf q})
 ({\bf p}-{\bf K})\cdot\nabla_{\bf K} R({\bf K},{\bf q})\bigg\},
\\ &&
= e^{i{\bf K}\cdot{\bf v}}\bigg\{
g({\bf x},{\bf x}') F\big({\bf K},{\bf u}\big)
-i\big(\nabla_{\bf v} g({\bf x},{\bf x}') \big)\cdot
\nabla_{\bf K}F\big({\bf K},{\bf u}\big)\bigg\}
\nonumber \\
\end{eqnarray}
where in this case
\begin{eqnarray}\Label{AppB.4}
 g({\bf x},{\bf x}') &=& 
{\sin\Delta_x(x-x')\over\pi(x-x')}
{\sin\Delta_y(y-y')\over\pi(y-y')}
{\sin\Delta_z(z-z')\over\pi (z.-z')},
\nonumber \\
\end{eqnarray}
as in QKI.

Using the formula (\ref{AppB.3}), tjhe commutation relations 
(\ref{in.wv10},\ref{in.wv10}) and 
the Gaussian factorization of the correlation functions, it is not difficult to 
derive the form of the correction (\ref{le302}).  It should be noted that the 
smoothness assumption will not be valid at the edge of the non-condensate band, 
where a boundary layer will form.  This  results from the fact that the 
Liouvillian straming in the time-dependent potential is not parallel to the 
actual fixed boundary between the two bands.

\EndTwoColumn
\end{document}